\documentclass[preprint,12pt]{elsarticle}

\usepackage{algorithm2e}
\usepackage{amsmath}
\usepackage{amssymb}
\usepackage{braket}
\usepackage{tikz}
\usetikzlibrary{quantikz2}
\usepackage[noabbrev]{cleveref}

\usepackage{subcaption}
\usepackage{graphicx}
\usepackage{tikz}
\usepackage{xcolor}

\definecolor{matplotlibgreen}{RGB}{0, 153, 0}
\definecolor{matplotlibblue}{RGB}{0, 102, 204} 
\definecolor{matplotlibgray}{RGB}{128, 128, 128} 
\definecolor{matplotlibblack}{RGB}{0, 0, 0} 
\definecolor{matplotliborange}{RGB}{255, 128, 0} 

\DeclareRobustCommand\fullblack  {\tikz[baseline=-0.6ex]\draw[thick, matplotlibblack] (0,0)--(0.4,0);}
\DeclareRobustCommand\dashedgreen{\tikz[baseline=-0.6ex]\draw[thick, matplotlibgreen, dashed] (0,0)--(0.4,0);}
\DeclareRobustCommand\fullblue{\tikz[baseline=-0.6ex]\draw[thick, matplotlibblue] (0,0)--(0.4,0);}
\DeclareRobustCommand\dashedorange{\tikz[baseline=-0.6ex]\draw[thick, matplotliborange, dashed] (0,0)--(0.4,0);}

\DeclareRobustCommand\dotdashblue{\tikz[baseline=-0.6ex]\draw[thick, matplotlibblue, dash dot] (0,0)--(0.4,0);}
\DeclareRobustCommand\dottedgray {\tikz[baseline=-0.6ex]\draw[thick, matplotlibgray, dotted] (0,0)--(0.4,0);}

\makeatletter
\def\ps@pprintTitle{%
    \let\@oddhead\@empty
    \let\@evenhead\@empty
    \def\@oddfoot{\hfill \footnotesize \today}%
    \let\@evenfoot\@oddfoot}
\makeatother
\begin{document}

\begin{frontmatter}

\title{Dynamic Circuits for the Quantum Lattice-Boltzmann Method}
\cortext[author] {Corresponding author.\\\textit{E-mail address:} david.wawrzyniak@tum.de}
\author[a]{David Wawrzyniak\corref{author}}
\author[b]{Josef Winter}
\author[b]{Steffen Schmidt}
\author[b]{Thomas Indinger}
\author[c]{Christian F. Janßen}
\author[c]{Uwe Schramm}
\author[a,b]{Nikolaus A. Adams}

\affiliation[a]{organization={Technical University Munich, Munich Institute of Integrated Materials, Energy and Process Engineering},
            addressline={Lichtenbergstr. 4a}, 
            city={Garching},
            postcode={85748}, 
            state={Bavaria},
            country={Germany}}

\affiliation[b]{organization={Technical University of Munich, School of Engineering \& Design, Department of Engineering Physics and Computation, Chair of Aerodynamics},
            addressline={Boltzmannstr. 15}, 
            city={Garching},
            postcode={85748}, 
            state={Bavaria},
            country={Germany}}

\affiliation[c]{organization={Altair Engineering Inc.},
            addressline={1820 E. Big Beaver Road}, 
            city={Troy},
            postcode={48083}, 
            state={Michigan},
            country={United States}}

\begin{abstract}
We propose a quantum algorithm for the linear advection-diffusion equation (ADE) Lattice-Boltzmann method (LBM) that leverages dynamic circuits. 
Dynamic quantum circuits allow for an optimized collision-operator quantum algorithm, introducing partial measurements as an integral step. 
Efficient adaptation of the quantum circuit during execution based on digital information obtained through mid-circuit measurements is achieved.

The proposed new collision algorithm is implemented as a fully unitary operator, which facilitates the computation of multiple time steps without state reinitialization. 
Unlike previous quantum collision operators that rely on linear combinations of unitaries, the proposed algorithm does not exhibit a probabilistic failure rate. 
Moreover, additional qubits no longer depend on the chosen velocity set, which reduces both qubit overhead and circuit complexity.

Validation of the quantum collision algorithm is performed by comparing results with digital LBM in one and two dimensions, demonstrating excellent agreement. 
Performance analysis for multiple time steps highlights advantages compared to previous methods.

As an additional variant, a hybrid quantum-digital approach is proposed, which reduces the number of mid-circuit measurements, therefore improving the efficiency of the quantum collision algorithm.

\end{abstract}




\end{frontmatter}

\section{Introduction}

There has been a growing interest in quantum hardware and quantum algorithms in recent years. 
By exploiting exponential scaling and specific quantum effects such as entanglement and interference \citep{nielsen2010quantum}, quantum algorithms have the potential to accelerate simulation algorithms or enable simulations that are otherwise impractical with conventional approaches. 
Computational fluid dynamics (CFD) have received significant attention despite computational challenges, as the advantages of quantum computing have disruptive potential.
Multiple approaches are being investigated to incorporate quantum computing into CFD.

A typical approach is using the Harrow-Hassidim-Lloyd (HHL) algorithm \citep{harrow2009quantum} to solve linear systems of equations, with exponential speed-up over classical algorithms.
The application of the HHL algorithm to CFD has been demonstrated in various studies \citep{ye2024hybrid,gopalakrishnan2024solving,lapworth2022hybrid}.
Oz et al. \citep{oz2022solving} presented a quantum algorithm to solve the Burgers equation.
An alternative method for addressing fluid dynamics problems involves the use of variational quantum algorithms, which have been successfully applied in various instances \citep{liu2024variational, jaksch2023variational, ingelmann2024two, pfeffer2022hybrid}.

An important approach for solving NSE using quantum algorithms is the Lattice-Boltzmann method (LBM) \citep{lallemand2021lattice}.
The LBM uses latent encodings of discrete velocity sets to approximate single-particle distribution functions. 
A low Mach-number approximation of those is achieved by multiscale expansion, where in its simplest form, the relaxation term is based on the Bhatnagar–Gross–Krook (BGK) approximation.
The relaxation term introduces local non-equilibrium effects and thus nonlinearity.
The first quantum algorithms were proposed for lattice gas automata \citep{boghosian1998quantum,meyer1996quantum,meyer2002quantum,yepez2001quantum,yepez2001type}, the precursor of LBM.
These approaches have recently experienced renewed interest, as shown by recent advancements \citep{zamora2025efficient}.
Mezzacapo et al. \citep{mezzacapo2015quantum} presented the first quantum simulator based on lattice kinetic formalism for simulating flow phenomena.
Quantum algorithms for the collisionless Boltzmann equation have been proposed by Todorova and Steijl \citep{todorova2020quantum} and by Schalkers and Möller \citep{schalkers2022efficient}.
Budinksi \citep{budinski2021quantum,ljubomir2022quantum} developed the first quantum algorithm to solve the linear-advection diffusion equation using the Lattice-Boltzmann method.
Sanavio et al. \citep{sanavio2024lattice} proposed a Carleman linearization for the non-linear collision step of the LBM.
While the approximation is constrained to moderate Reynolds numbers, the number of variables grows exponentially with increasing expansion order. 
The feasibility of this approach remains an area for further exploration.

This publication proposes a quantum algorithm for the Lattice-Boltzmann Method that incorporates a linearized collision operator under the assumption that $\Delta t / \tau = 1$. 
The classical equilibrium distribution function is reformulated to facilitate quantum computation while preserving equivalence to digital computation after measurements.

The proposed reformulated collision operator offers several advantages. 
It can be expressed as unitary operator constructed using standard RY gates, eliminating the need for the linear combination of unitaries approach employed in previous implementations. 
This ensures application without failure rate, enabling predictable time stepping within the algorithm. 
Moreover, the algorithm leverages dynamic circuits, which reduce the computational complexity by incorporating mid-circuit measurements to adapt subsequent operations based on measurement outcomes. 
An alternative version of the algorithm employs classical pre-processing, which minimizes the reliance on mid-circuit measurements in the dynamic circuit setup.
The proposed algorithm is validated by comparing it with a digital LBM solver.

The remainder of the paper is organized as follows.
\Cref{Sec.2} introduces the LBM equations and describes the modifications implemented for the quantum algorithm. 
\Cref{sec.3} details the quantum algorithm, including its extension to arbitrary dimensions and velocity sets and the optimization strategies for reducing mid-circuit measurements through classical pre-processing. 
\Cref{sec.4} presents validation results using noise-free, sampling-based quantum simulators and a digital LBM algorithm. 
Finally, \cref{sec.5} presents the conclusions of this study.

\section{Methodology}\label{Sec.2}
\subsection{Lattice-Boltzmann method}
The time evolution equation for discrete distribution functions $f_i(\mathbf{x},t)$ in the Lattice-Boltzmann method is given by
\begin{equation}
 f_i(\mathbf{x}+\mathbf{c}_i\Delta t, t+\Delta t) - f_i(\mathbf{x},t) =   \Omega_i\left(f\right),
\end{equation}
where $\Omega_i\left(f\right)$ is the collision operator, $\Delta t$ is the time step size, and $c_i$ is the microscopic velocity. 
The opposing direction of a distribution function $f_i$ is denoted by $f_{\bar{i}}$ with the corresponding microscopic velocity $c_{\bar{i}}=-c_i$.
The number of discrete velocities is determined by the choice of velocity set.
We employ the Bhatnagar-Gross-Krook (BGK) collision operator \citep{bhatnagar1954model} with a single relaxation time.
The Lattice-Boltzmann equation then becomes \citep{kruger2017lattice,mohamad2011lattice,chen1998lattice,PhysRevLett.61.2332,higuera1989lattice,chen1992recovery}
\begin{equation}
 f_i\left(\mathbf{x}+\mathbf{c}_i\Delta t,t+\Delta t\right)=\left(1-\frac{\Delta t}{\tau}\right)f_i(\mathbf{x},t)+\frac{\Delta t}{\tau} f_i^{eq}(\mathbf{x},t),
    \label{eq.lbm}
\end{equation}
where $\tau$ is the relaxation time, and $f_i^{eq}(\mathbf{x},t)$ is the equilibrium distribution function.
Furthermore, assuming $\Delta t / \tau = 1$ \citep{junk1999new}, \cref{eq.lbm} simplifies to 
\begin{equation}
 f_i\left(\mathbf{x}+\mathbf{c}_i\Delta t,t+\Delta t\right)=f_i^{eq}(\mathbf{x},t).
\end{equation}
The linearized equilibrium distribution function for the advection-diffusion equation (ADE) reads
\begin{equation}
 f_i^{eq}=w_i{\rho}\left(1+\frac{\mathbf{c}_i\cdot\mathbf{u}(\mathbf{x},t)}{c_s^2}\right),
    \label{eq.5}
\end{equation}
with $w_i$ being a weighting factor, $\rho$ being density, $\mathbf{c}_i$ represents the microscopic velocities, $c_s$ being a lattice constant often referred to as lattice speed of sound, and $\mathbf{u}(\mathbf{x},t)$ being the macroscopic velocity.
Macroscopic quantities, such as the density $\rho$, are obtained by calculating the moments of the distribution functions \citep{he1997priori,he1997theory,abe1997derivation}. 
Specifically, the density $\rho$ is computed using the zeroth moment
\begin{equation}
    \rho = \sum_i f_i.
\end{equation}

\subsection{Quantum-biased LBM collision operator}\label{sec.2.2}

In this section, we derive a quantum-biased collision operator for modified LBM using the example of the advection-diffusion equation (ADE).
We employ amplitude encoding to represent the distribution functions as a quantum state vector.
Modifications of the digital LBM collision algorithm are introduced to enable efficient quantum computing.
For a digital simulation, the collision step typically computes \cref{eq.5} under given assumptions. 
However, with the modified quantum-biased algorithm, rather the square root of \cref{eq.5} is computed as
\begin{equation}
 \sqrt{f_i^{eq}}=\sqrt{w_i{\rho}\left(1+\frac{\mathbf{c}_i\cdot\mathbf{u}(\mathbf{x},t)}{c_s^2}\right)}.
\end{equation}
Upon successful quantum computation of the collision step, the solution is embedded into the amplitudes of the quantum state vector \(\ket{\Psi}\), represented as
\begin{equation}
 \ket{\Psi} = \sum_k\sqrt{f_{i,k}^{eq}}\ket{k},
\end{equation}
where $k$ is a label for the states in the computational basis
For simplicity, we consider only the i-th velocity population.
A full state measurement directly yields the normalized $f_{0}$ as a result, only requiring a renormalization.

Previous approaches have implemented the collision step using a linear combination of unitaries (LCU) algorithm \cite{budinski2021quantum,ljubomir2022quantum,wawrzyniak2025quantum}. 
The non-unitary collision operator that computes \cref{eq.5} is represented as a sum of unitaries, employing ancillary qubits for implementation as a quantum circuit.
A significant drawback of this approach is the probabilistic representation of the collision operator. 
The operator is applied with a specific success rate, depending on the measurement of the ancillary qubit. 
If the measurement of the ancillary qubit indicates a failed application, the results are discarded, and a repetition of the entire quantum LBM iteration is necessary. 
For multiple time steps, the repeated application of the LCU-based collision operator may lead to exponential complexity.

\begin{figure}
    \centering
    \includegraphics[width=0.5\textwidth]{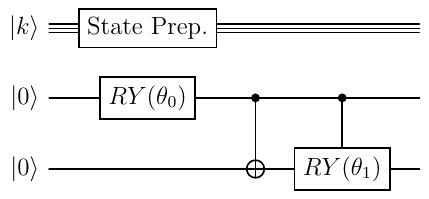}
    \caption{Quantum circuit for the collision step for a D1Q3 LBM.}
    \label{fig:qc_collision_intro}
\end{figure}

We prepare an alternative algorithm for the collision step that is not based on the LCU method and is implemented by standard unitary \texttt{RY}-gates. 
As a result, the algorithm does not require additional ancillary qubits and achieves a $100\%$ success rate in applying the collision operator. 
We will now discuss the modified collision process for simplicity using the D1Q3 velocity set. 
The D1Q3 velocity set emplys three discrete velocities, $c_0=0$, $c_1=1$, and $c_2=-1$ and corresponding weights, $w_0=2/3$, $w_1=1/6$, and $w_2=1/6$.
The equilibrium populations are given by
\begin{equation}
 \sqrt{f_0}=\sqrt{w_0} \sqrt{\rho},
    \label{eq.w_0}
\end{equation}
\begin{equation}
 \sqrt{f_1}=\sqrt{w_1} \sqrt{1+\frac{u}{c_s^2}} \sqrt{\rho}, \text{ and}
    \label{eq.f_1_eq}
\end{equation}
\begin{equation}
 \sqrt{f_2}=\sqrt{w_2} \sqrt{1-\frac{u}{c_s^2}} \sqrt{\rho},
    \label{eq.f_2_eq}
\end{equation}
where the weighting factors are related by $w_0+w_1+w_2=1$.
When preparing the state vector using the normalized $\sqrt{\rho_k}$, the state vector encoding the post-collision state, assuming a uniform and constant advection velocity, reads
\begin{multline}
 \ket{\Psi}=\sqrt{w_0} \sqrt{\rho_k}\ket{00}_Q\ket{k}+\sqrt{w_1} \sqrt{1+\frac{{u}}{c_s^2}} \sqrt{\rho_k}\ket{01}_Q\ket{k}\\+\sqrt{w_2} \sqrt{1-\frac{{u}}{c_s^2}} \sqrt{\rho_k}\ket{10}_Q\ket{k},
    \label{eq.10}
\end{multline}
under the constraint $(u/c_s^2)\le 1$.
It can easily be verified that \cref{eq.10} satisfies the normalization constraint by calculating the inner product $\braket{\Psi | \Psi}$, confirming that it is a valid quantum state vector.
The construction of the state vector in \cref{eq.10}, follows \citep{Wawrzyniak2024}, the quantum circuit implementing the collision is illustrated in \cref{fig:qc_collision_intro}.
The \texttt{RY}-gate is defined as
\begin{equation}\label{eq.ry}
 \texttt{RY} =     \begin{bmatrix}
 \cos\left(\theta/2\right) & -\sin\left(\theta/2\right) \\
 \sin\left(\theta/2\right) & \cos\left(\theta/2\right)
    \end{bmatrix}. 
\end{equation}
We define the arguments
\begin{align}
    \theta_0 = 2\arccos\left(\sqrt{w_0}\right) \label{eq.theta0},\text{ and}\\
    \theta_1 = 2\arccos\left(\sqrt{\frac{1}{2}\left(1+\frac{u}{c_s^2}\right)}\right)\label{eq.theta1}
\end{align}
for the \texttt{RY}-gates, where \cref{eq.theta1} is derived under the assumption of a constant and uniform macroscopic velocity.
Note that $\theta_0$, and $\theta_1$ directly follow from \cref{eq.w_0,eq.f_1_eq}, respectively.
In \cref{sec.3}, we present the corresponding quantum algorithm that accommodates general advection velocities.
Inserting \cref{eq.theta0,eq.theta1} into \cref{eq.ry} yields the matrices
\begin{align}
 \texttt{RY}(\theta_0) =\begin{bmatrix}
 \sqrt{w_0} & -\sqrt{w_{12}} \\
 \sqrt{w_{12}} & \sqrt{w_0}
    \end{bmatrix}, \\
 \texttt{RY}(\theta_1) =\begin{bmatrix}
 \sqrt{\frac{1}{2}\left(1+\frac{u}{c_s^2}\right)} & -\sqrt{\frac{1}{2}\left(1-\frac{u}{c_s^2}\right)} \\
 \sqrt{\frac{1}{2}\left(1-\frac{u}{c_s^2}\right)} & \sqrt{\frac{1}{2}\left(1+\frac{u}{c_s^2}\right)}
    \end{bmatrix},\label{eq.RY_gate}
\end{align}
where we use the identity $\sin{\left(\arccos{x}\right)}=\sqrt{1-x^2}$, and introduce $w_{12}=2w_1=2w_2$.
\Cref{eq.RY_gate} shows that $\texttt{RY}(\theta_1)$ contains the necessary information to compute $f_1$ and $f_2$ in the first column. 
This implies that the \texttt{RY}-gate simultaneously computes the information for both distribution functions $f_1$ and $f_2$ within one operation.

The quantum circuit in \cref{fig:qc_collision_intro} is constructed as follows.
First, we prepare the normalized quantum state vector $\ket{\Psi_0} = \sum_k \sqrt{\rho_k}\ket{00}_f\ket{k}$.
Applying a \texttt{RY}-gate with $\theta_0$ on the first qubit in the 'f' register results in
\begin{equation}
 \ket{\Psi_1}=\texttt{RY}(\theta_0)\ket{\Psi_0}=\sum_{k} \sqrt{w_0}\sqrt{\rho_k}\ket{00}_f\ket{k}+\sqrt{w_{12}}\sqrt{\rho_k}\ket{01}_f\ket{k}.
\end{equation}
Next, using a \texttt{CNOT} gate, the states are rearranged 
\begin{equation}
 \ket{\Psi_2}=CX\ket{\Psi_1}=\sum_{k} \sqrt{w_0}\sqrt{\rho_k}\ket{00}_f\ket{k}+\sqrt{w_{12}}\sqrt{\rho_k}\ket{10}_f\ket{k}.
\end{equation}
After using the last controlled-\texttt{RY} gate, the state vector becomes
\begin{multline}
 \ket{\Psi_3}=\texttt{CRY}(\theta_1)\ket{\Psi_2}=\sum \sqrt{w_0\rho_k}\ket{00}_f\ket{k}+\\\sqrt{w_1\left(1+\frac{u}{c_s^2}\right)\rho_k}\ket{10}_f\ket{k}+\sqrt{w_2\left(1-\frac{u}{c_s^2}\right)\rho_k}\ket{11}_f\ket{k},
\end{multline}
which corresponds to \cref{eq.10}, except for the state labels. 
These labels can be easily adjusted using additional \texttt{CNOT}-gates.

\section{Quantum algorithm}\label{sec.3}

Based on the proposed unitary collision, we formulate a quantum algorithm for the BGK Lattice-Boltzmann discretization of the advection-diffusion equation.
It enables the computation of multiple time steps before a full state measurement is performed. 
Incorporating dynamic circuits reduces the required number of gates and circuit depth.
Its structure is divided into four blocks: (i) initialization, (ii) collision, and (iii) streaming. 
One copy of the macroscopic scalar field is encoded in the initialization block in the state vector.
In the collision step, the linearized equilibrium distribution function assuming $\Delta t/\tau=1$ is computed using dynamic circuits and the previously introduced methodology.
Established quantum algorithms are used for the streaming step.
After the computation of a selected number of time steps, a full state measurement is performed, which intrinsically sums the distribution functions to compute the macroscopic variable.

\subsection{Initialization and state preparation}\label{sec.3.1}

The quantum circuit required for the algorithm only depends on the spatial discretization of the problem to be solved.
The number of qubits $n_q$ to encode the domain is
\begin{equation}
 n_q=\log_2(N_xN_yN_z),
\end{equation}
where $N_x$, $N_y$, and $N_z$ are the number of lattice cells in the respective spatial dimension.
For simplicity, we restrict the number of lattice cells in every spatial dimension to a power of two.
The total number of qubits $n_t$ required for the algorithm is
\begin{equation}
 n_t=n_q+1,
\end{equation}
with one extra ancillary qubit independently of the underlying velocity set, unlike previous work \citep{Wawrzyniak2024,budinski2021quantum,ljubomir2022quantum}.
The state vector is prepared using standard amplitude encoding \cite{1629135}. 
First, d-dimensional initial condition $(d=1,2,3)$ of density $\rho$ is flattened into a vector, followed by taking the square root of each entry in the density vector.
Subsequently, it is normalized and encoded into the state vector $\ket{\Psi}$
\begin{equation}
 \ket{\Psi} = \sum_k^{N_q} \sqrt{\rho_k}\ket{k}\ket{0}_a.
    \label{eq.init}
\end{equation}
Therein, $k$ is a label for the states in the computational basis, $N_q=2^{n_q}$ is the total number of possible states for the qubits $n_q$, and $\ket{0}_a$ is the state of the ancillary qubit denoted with a subscript 'a'.
Furthermore, we assume a normalized state.
The ancillary qubit is the least significant in the quantum register. 
It should be noted that this qubit remains unaltered during the initialization process.

\subsection{Collision and streaming block in dynamic circuits}\label{sec.3.2}

In this section, we develop a collision and streaming block for the quantum algorithm of the Lattice Boltzmann Method using dynamic circuits. 
It builds upon the collision step described in \cref{sec.2.2}, with adjustments to accommodate general advection velocities and the introduction of dynamic circuits to reduce the gate count and depth of the algorithm.

\begin{figure}
    \centering
    \begin{subfigure}{0.5\textwidth}
        \includegraphics[width=\textwidth]{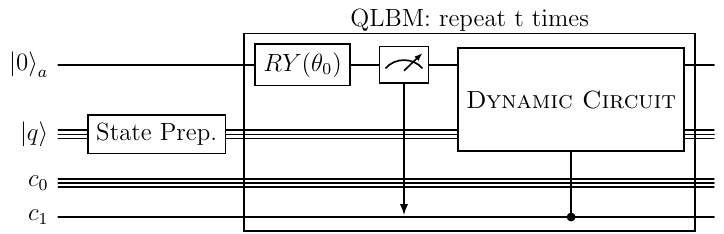}
        \caption{The general quantum circuit for the QLBM.}
        \label{fig:full_circuit}
    \end{subfigure}
    \hfill
    \begin{subfigure}{0.4\textwidth}
        \includegraphics[width=\textwidth]{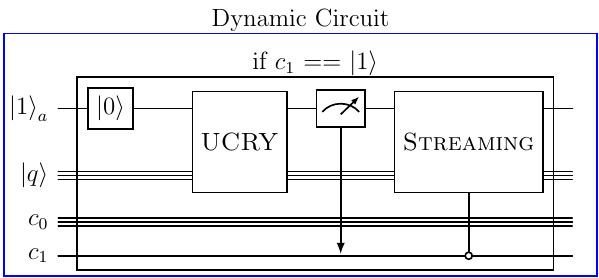}
        \caption{The dynamic circuit within QLBM.}
        \label{fig:dynamic_circuit}
    \end{subfigure}
    \hfill
    \begin{subfigure}{0.6\textwidth}
        \includegraphics[width=\textwidth]{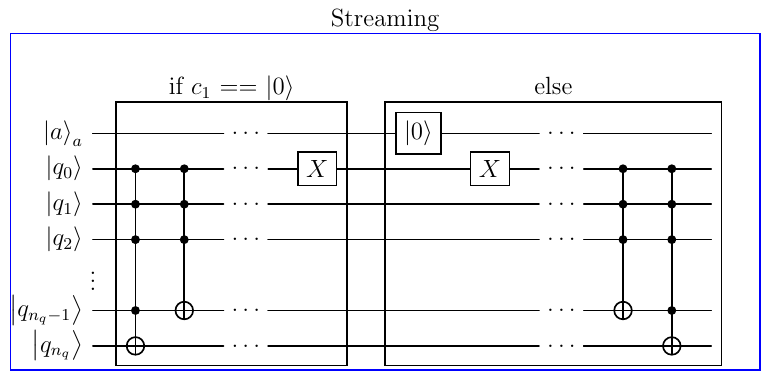}
        \caption{Streaming circuit within QLBM.}
        \label{fig:streaming}
    \end{subfigure}
    \caption{All quantum circuits for the dynamic circuit implementation of the D1Q3 QLBM.}
    \label{fig:all_circuits}
\end{figure}

Dynamic circuits are quantum circuits that incorporate classical processing during their execution. 
This allows for actions based on mid-circuit measurements like "if-else" logic in digital computing. 
The quantum algorithm can adaptively modify its behavior based on mid-circuit measurement outcomes.
The quantum circuits corresponding to the D1Q3 velocity set are sketched in \cref{fig:all_circuits}. 
Generalizing this algorithm for two and three spatial dimensions is straightforward and will be discussed in the subsequent sections. 
The quantum circuit for a two-dimensional D2Q9 velocity set is provided in the \cref{App.1}.
The collision block of the algorithm can be divided into two main algorithmic steps. 
The first step involves a probabilistic process, determining which equilibrium distribution functions will be computed. 
This is depicted in \cref{fig:full_circuit}, where the first \texttt{RY}-gate is applied on the ancillary qubit, followed by a mid-circuit measurement determining the subsequent computations. 
This measurement enables dynamic conditioning within the algorithm.
The state vector before measurement is given by
\begin{equation}
 \ket{\Psi} = \sum_k^{N_q} \sqrt{\rho_k}\ket{k}\sqrt{w_0}\ket{0}_a+\sqrt{\rho_k}\ket{k}\sqrt{w_{12}}\ket{1}_a,
\end{equation}
where \cref{eq.w_0} is applied for the \texttt{RY}-gate. 
After the measurement of the ancilla state, the classical information is processed digitally to determine the next steps of the algorithm.
Suppose the ancilla qubit is measured in the $\ket{0}_a$ state, corresponding to the computation of the zeroth distribution function. In that case, the calculation of this time step is completed, as no streaming is required for the resting population, and the next time step can be immediately computed.
If $\ket{q}_a=\ket{1}$ is measured and $\ket{q}_a$ is reset back to the state $\ket{0}$, the state vector is renormalized to \cref{eq.init}.
In the following, we change the notation and absorb the ancillary qubit $\ket{q}_a$ into $\ket{k}$.
As a result, it becomes evident that only every second entry in the state vector is occupied since the ancillary qubit is in the classical state $\ket{0}_a$. 
The state vector reads
\begin{equation}
 \ket{\Psi} = \sum_{j=0}^{N_\text{total}-1} \sqrt{\rho_j} \ket{2j},
\end{equation}
where the new notation \(\ket{j}\) represents all combined states of the ancilla and \(\ket{k}\), while \(\ket{2j}\) indicates that only every second state is populated, and $N_\text{total}=2^{n_q+1}$.
Uniformly controlled Pauli-Y rotations, with the unitary
\begin{equation}
 \texttt{UCRY} =
    \begin{bmatrix}
 RY(\theta_{1,0}) & 0 & 0 & \cdots & 0 \\
    0 & RY(\theta_{1,1}) & 0 & \cdots & 0 \\
    0 & 0 & RY(\theta_{1,2}) & \cdots & 0 \\
    \vdots & \vdots & \vdots & \ddots & \vdots \\
    0 & 0 & 0 & \cdots & RY(\theta_{1,n})
    \end{bmatrix}\label{eq.25}
\end{equation}
are then applied to the state vector. In each \cref{eq.25}, each $\texttt{RY}(\theta)$ is a 2x2 unitary matrix that applies collision to each lattice site, computing a distribution function, and to its opposite oriented spatial direction in parallel.
This evolves the state vector to 
\begin{multline}
 \ket{\Psi} = \sum_{j=0}^{N_\text{total}-1} \sqrt{\frac{1}{2}\left(1 + \frac{u_{j}}{c_s^2}\right)} \sqrt{\rho_{j}} \ket{2j} \\ +\sum_{j=0}^{N_\text{total}-1} \sqrt{\frac{1}{2}\left(1 - \frac{u_{j}}{c_s^2}\right)} \sqrt{\rho_{j}} \ket{2j+1}.
    \label{eq.aftercol}
\end{multline}
Even-indexed states in $\ket{j}$ correspond to the direction $c_1$, while odd-indexed states correspond to the opposing direction $c_2$. 
Each component of the state vector now encodes the equilibrium populations for both directions. 
We emphasize that computing $\theta_{1,n}$ requires determining only a single direction during the classical preprocessing step. 
The amplitudes of the state vector in \cref{eq.aftercol} do not numerically match the respective equilibrium distribution function data in equation (\ref{eq.f_1_eq}) and (\ref{eq.f_2_eq}). 
This is because the initial measurement step, which determines whether the ancilla qubit is in \(\ket{0}_a\) or \(\ket{1}_a\), is not yet reflected in the state vector equation due to the renormalization after measurement. 
However, the state in \cref{eq.aftercol} occurs with a probability  $p = 2w_1 = 2w_2$, ensuring that evaluated probabilities agree with the equilibrium distribution functions after associated measurements.
The ancilla qubit is measured again, determining whether the state is in \(\ket{0}_a\), corresponding to $f_1^{eq}$, or \(\ket{1}_a\), corresponding to $f_2^{eq}$. 
Based on this measurement, the corresponding streaming circuit, as depicted in \cref{fig:streaming}, is applied. 
The time step is completed by propagating the distribution function along their respective velocity directions. 
Depending on the mid-circuit measurement outcomes, either $f_0$ or the streamed distribution functions $f_1$ or $f_2$ are encoded in the state vector.
Repeating the dynamic collision and streaming circuit allows for the computation of multiple time steps before a full state measurement.

\subsection{Time marching in the QLBM}\label{sec.3.3}

Consecutive time steps can be computed using the above streaming and collision block without requiring a full state measurement or reinitialization \citep{Wawrzyniak2024}. 
The block of \texttt{RY}-gate, measurement process, and the dynamic circuit is iterated until a final time is reached.
We consider the linear operator $O_{LBM}$, which propagates the LBM density of by one time step
\begin{equation}
    \rho(\mathbf{x},t+1) = O_{LBM}\rho(\mathbf{x},t).
    \label{eq.operator_lbm}
\end{equation}
The density $\rho(\mathbf{x},t)$ is the zeroth moment of the distribution function. 
For simplicity, we ascribe the D1Q3 velocity set with $\Delta t / \tau =1$.
Therefore, \cref{eq.operator_lbm} can be written as
\begin{equation}
    \rho(\mathbf{x},t+1) = O_{LBM}\left(f_0(\mathbf{x},t)+f_1(\mathbf{x},t)+f_2(\mathbf{x},t)\right).
\end{equation}
Due to linearity $O_{LBM}$ can be applied to each distribution separately
\begin{equation}
    \rho(\mathbf{x},t+1) = O_{LBM}f_0(\mathbf{x},t)+ O_{LBM}f_1(\mathbf{x},t)+ O_{LBM}f_2(\mathbf{x},t).
\end{equation}
Thus, the next time step can be computed by applying $O_{LBM}$ to each distribution function, followed by summation.
Therefore, evolving the distribution functions in time allows for the computation of multiple time steps. 
Upon application of the quantum circuit for $t$ iterations, the entire state is measured and stored under the same label in  $\ket{0}_a$, which intrinsically sums up the distribution functions and computes the density distribution after $t$ time steps.
Except for renormalization, no separate digital post-processing is necessary.

\subsection{Extension to arbitrary spatial dimensions and velocity sets} \label{sec.3.4}

The extension for arbitrary spatial dimensions and velocity sets is straightforward and can be done using slight modifications of the proposed building blocks in \cref{sec.3.1,sec.3.2}.
Velocity sets with multiple discrete velocities differ only in the \texttt{RY}-gate and measurement process, where the specific velocity direction to be computed is chosen. 
We now discuss the extension to general velocity sets. 
Starting with an initialized state vector, we require only one ancillary qubit
\begin{equation}
 \ket{\Psi} = \sum_k \sqrt{\rho_k}\ket{0}_a\ket{k},
    \label{eq.init2}
\end{equation}
where the state is assumed to be normalized. 
We define the quantum density operator $\hat{\sigma}$ as
\begin{equation}
 \hat{\sigma} = \sum_j p_j\ket{\Psi_j}\bra{\Psi_j},
\end{equation}
where $\hat{\sigma}$ describes an ensemble of states, and the system is in state $\ket{\Psi_j} \bra{\Psi_j}$ with probability $p_j$.
Each velocity set has an associated weight $w_j$ to its corresponding distribution functions $f_j$.
We set the $p_j=w_j$, so that each subsystem $\ket{\Psi_j} \bra{\Psi_j}$ corresponds to one distribution function.
Applying subsequent \texttt{RY}-gates, along with tracing out and resetting the ancillary qubit, the quantum system is projected onto one of the subsystems. It evolves under the corresponding unitary operators that compute the associated distribution functions.

\subsection{Hybrid classical-quantum algorithm}

The proposed quantum algorithm can be further optimized by hybridization with a classical preprocessing step, leading to a reduction in the number of mid-circuit measurements. 
As described in \cref{sec.3.4}, specific distribution functions are identified by \texttt{RY}-gates followed by quantum measurements. 
Certain computations can be performed by a classical preprocessing step.
Initially, a list is generated where each population $f_i$ is paired with its corresponding weight $w_i$. 
Opposing distribution functions $f_i$ and $f_{\bar{i}}$ are combined into a single entry $f_{i,\bar{i}}$.
Their combined weights correspond to $w_{i,\bar{i}} = w_i + w_{\bar{i}}$. 
For example, for the D2Q9 velocity set, the list is structured as
\begin{equation}
 f = [f_0 : w_0, f_{1,3} : w_{1,3}, f_{2,4} : w_{2,4}, f_{5,7} : w_{5,7}, f_{6,8} : w_{6,8}]^T.
\end{equation}
During each time step $t$, a distribution function is selected based on a digital probabilistic sampling process, where the likelihood of selecting $f_i$ is proportional to its weight $w_i$
\begin{equation}
 p(f_i) = w_i.
\end{equation}
This sampling is repeated for the number of shots $s$ executed by the quantum algorithm, resulting in an instruction array with dimensions $[t, s]$. 
This array dictates the construction of the collision and streaming blocks for the quantum circuits.
For each shot, the circuit initialization remains the same. 
Instructions for $f_0$ are represented by an identity operator, while all other instructions correspond to specific dynamic circuit components, as detailed in \cref{fig:dynamic_circuit}. 
Consequently, efficient classical preprocessing reduces the number of required mid-circuit measurements and improves overall efficiency.

\subsection{Computational complexity}

In this section, we analyze the computational complexity of the proposed algorithm by evaluating the cost of each algorithmic block in terms of gate count. 
\subsubsection*{Initialization}
The complexity of the initialization block depends primarily on the initial condition to be encoded. 
For a uniform initial condition, the computational cost is minimal, requiring only one Hadamard gate per qubit, resulting in a total gate count of $n$, where $n$ is the number of qubits. 
However, preparing a general quantum state vector incurs exponentially cost, necessitating $2 \times 4^n - (2n+3) \times 2^n + 2n$ gates \cite{1629135}.
The number of qubits required for the initialization step in the proposed algorithm is determined solely by the spatial discretization of the computational domain and is given by $n = \log_2(N_x N_y N_z)$, where  $N_x$, $N_y$, and $N_z$ are the respective numbers of lattice cells in each spatial dimension. 
In contrast, in the prior algorithms \cite{budinski2021quantum}, the number is proportional to both the domain size and the velocity set, $\lceil \log_2 (q N_x N_y N_z) \rceil$, where $q$ denotes the number of discrete velocities in the velocity set.
General state preparation still incurs an exponential cost, albeit the exponent is reduced by a factor of q.
\subsubsection*{Collision}
The collision block is implemented as a probabilistic sequence, determining the specific population to be calculated. 
Uniformly controlled Y-rotation gates (\texttt{UCRY}) are applied to all populations except for $f_0$. 
The length of this probabilistic sequence depends on the number of distribution functions $q$ in the chosen velocity set and the measurement outcome. 
The sequence requires up to $(q-1)/2$ single-qubit Y-rotation gates and measurements, concluding when a measurement yields $\ket{0}$ state. 

This probabilistic sequence can be entirely omitted by employing the hybrid classical-quantum approach. 
The \texttt{UCRY} gates require $2^{n-1}$ controlled-NOT (\texttt{CNOT}) gates \cite{bergholm2005quantum} and are applied to $n_q+1$ qubits. 
Importantly, these gates are not applied when the measurement outcome corresponds to the computation of $f_0$. 
As a result, the average computational cost of the collision block is determined by the chosen velocity set and the associated weights. 
For instance, for the D1Q3 velocity set, the \texttt{UCRY} gate is applied in $1/3$ of all shots on average for one time step, whereas for the D2Q9 velocity set, the ratio is $5/9$.

Previous algorithms based on the Linear Combination of Unitaries (LCU) approach can be executed using a diagonal gate, which requires at most $2^n-2$ \texttt{CNOT} gates \cite{1629135} and is applied to $n_b+1$ qubits \cite{wawrzyniak2025quantum}. 
Although both approaches exhibit similar scaling, the proposed algorithm applies the collision block only to a subset of shots. It operates on a reduced number of qubits, thereby achieving an efficiency advantage.
\subsubsection*{Streaming}
The streaming operator in our implementation is constructed using a multi-controlled X-gate sequence. 
It can be reformulated based on the Quantum Fourier Transform (QFT) \cite{schalkers2022efficient} as an alternative to the specific approach described. 
The decomposition of these gates into native quantum gates remains an active area of research, and its application to the QLBM has been further discussed by Budinksi \cite{budinski2023efficient} and Schalkers et al. \cite{schalkers2022efficient}. 

Although the realization of the streaming operator is known from prior QLBM algorithms, the methodology proposed here provides some significant advantages. 
Specifically, we reduce complexity by applying the streaming operator on $\log_2 (N_x N_y N_z)$ qubits instead of $\lceil \log_2 (q N_x N_y N_z) \rceil+1$ \cite{budinski2021quantum}. 
Furthermore, only a single population needs to be streamed per time step, whereas previous methods required streaming operations for all populations,
\subsubsection*{Time stepping}
Time stepping does not require reinitialization of the quantum state and is realized by repeated collision and streaming operator applications. 
Consequently, the computational complexity for simulating multiple time steps scales linearly with the number of time steps.
Previous algorithms \cite{budinski2021quantum,ljubomir2022quantum,wawrzyniak2025quantum} rely on reinitialization of the full state vector during each time step.
\section{Results}\label{sec.4}

The quantum circuits are built and executed using the Qiskit SDK version 1.0.2.
A noise-free sampling-based simulator is employed to execute the quantum circuits.
We assume, for all cases, periodic boundary conditions.

\subsection{Validation}\label{sec.4.1}

\begin{figure}
    \centering
    \begin{subfigure}{0.45\textwidth}
        \includegraphics[width=\textwidth]{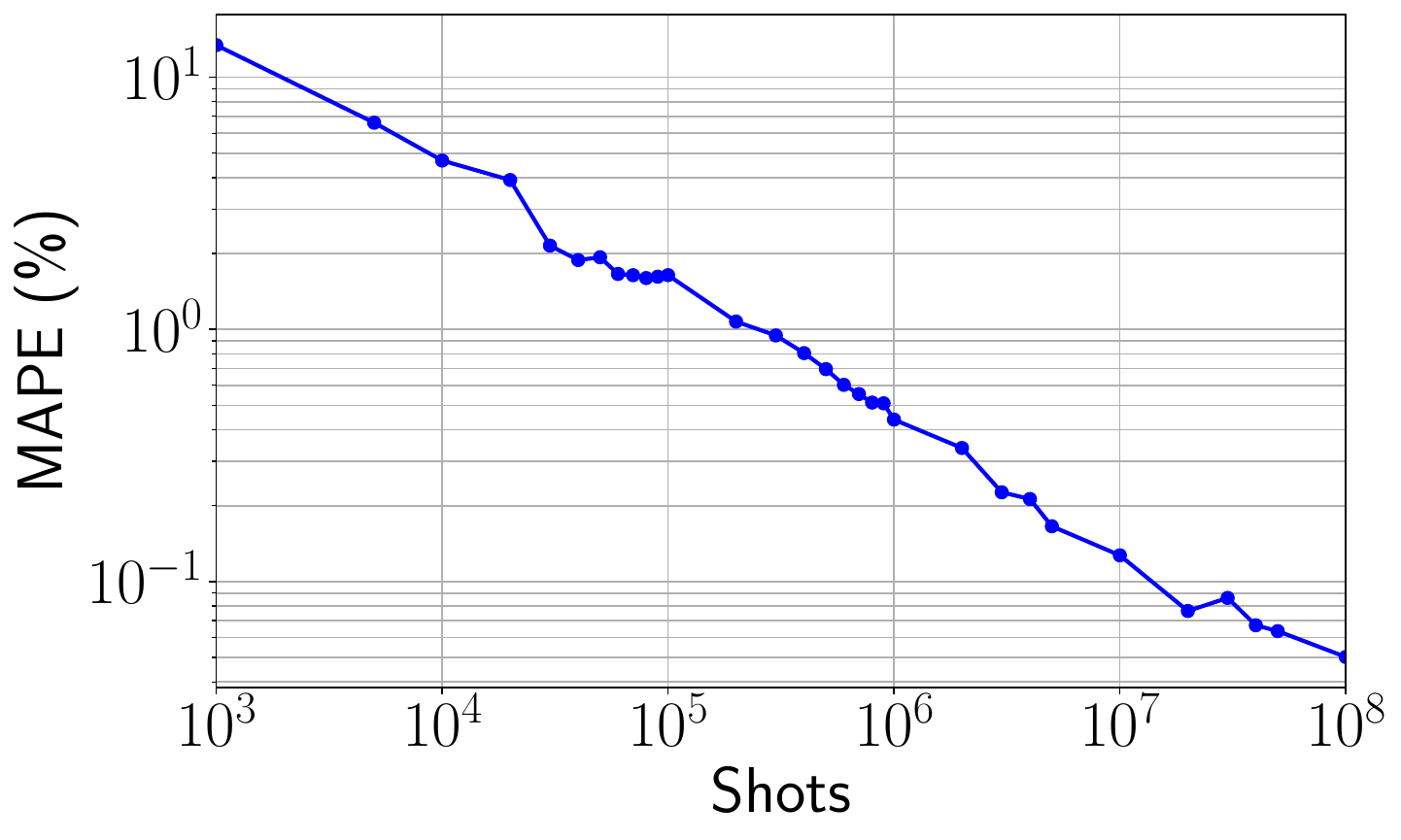}
        \caption{MAPE for QLBM and digital LBM using the D1Q3 velocity set with $N=32$ grid points.}
        \label{fig:MAPE_D1Q3}
    \end{subfigure}
    \hfill
    \begin{subfigure}{0.45\textwidth}
        \includegraphics[width=\textwidth]{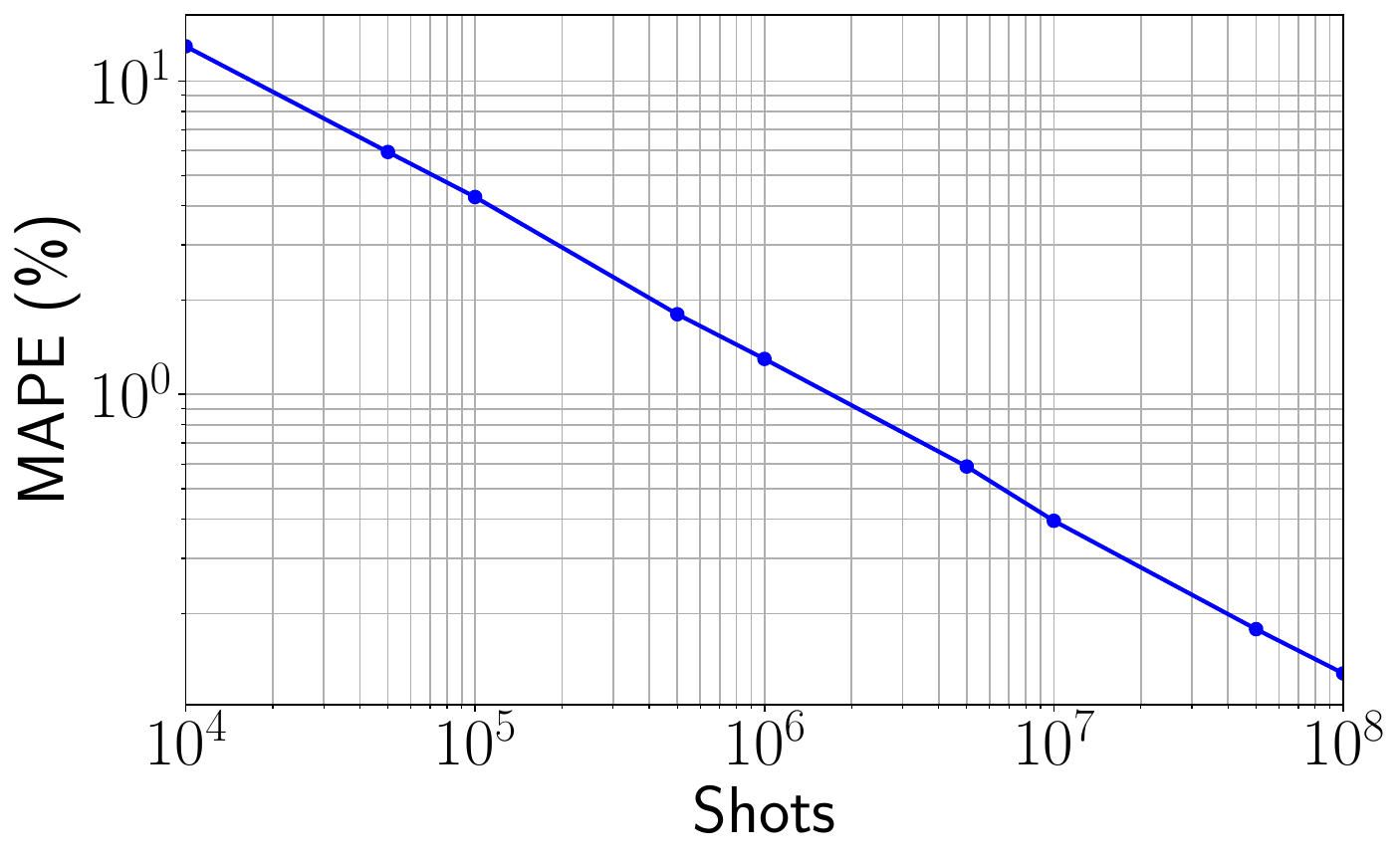}
        \caption{MAPE for QLBM and digital LBM using the D2Q9 velocity set with $[16 \times 16]$ grid points.}
        \label{fig:MAPE_2Q9}
    \end{subfigure}
    \caption{MAPE for the computation of one time step using QLBM and digital LBM using D1Q3 and D2Q9 velocity sets.}
    \label{fig:MAPEoneTimestep}
\end{figure}

We validate our quantum algorithm by comparing the results with a digital LBM algorithm.
The error is measured using the mean absolute percentage error (MAPE) defined as
\begin{equation}
 \text{MAPE} = 100\frac{1}{N} \sum_{i=1}^{N} \left| \frac{\rho_i^{\text{LBM}} - {\rho}_i^\text{QLBM}}{\rho_i^{LBM}} \right|,
\end{equation}
where $N$ is the number of grid points, $\rho^{LBM}$ is the reference result, and $\rho^{QLBM}$ is the result of the quantum algorithm. 
The domain size remains constant, while the number of shots of the quantum simulator varies.
The boxcar function gives the initial condition
\begin{equation}
    \rho(\mathbf{x},t=0) = 0.1 + 0.1 [H(\mathbf{x}-\mathbf{x}/2+3)-H(\mathbf{x}-\mathbf{x}/2-3)],
\end{equation}
where $H$ is the Heaviside step function, with the width of the rectangle being $6$ and its amplitude being $0.2$. 
The rectangle is symmetrically placed in the middle of the domain, and the surrounding density is 0.1.
The boundary conditions are periodic, and we assume a uniform velocity $u=0.1$ for the one-dimensional case and $u(x,y)=0.1$, $v(x,y)=0.1$ for the two-dimensional case.
The one-dimensional case is computed using $N=32$ grid points, and the two-dimensional case is computed using  $[16\times16]$ grid points.
One time step of the algorithm is computed.
The corresponding error plots are shown in \cref{fig:MAPE_D1Q3,fig:MAPE_2Q9} for the two cases, respectively.
The error decreases with an increasing number of shots, steadily approaching the reference result.
This is to be expected as the statistical noise decreases with increasing shot counts, and the quantum algorithm converges to the exact result \cite{wawrzyniak2025quantum}.

\newpage

\subsection{Multiple time steps}

In this section, we assess the performance of the quantum algorithm when applied to simulations over multiple time steps. 
As an initial validation, we consider the same test one-dimensional case as described in \cref{sec.4.1}, computing results for $t=[10, 50, 100, 150, 200, 250]$ time steps using $[10^5, 10^6, 10^7]$ shots.

\begin{figure}
    \centering
    \begin{subfigure}{0.45\textwidth}
        \includegraphics[width=\textwidth]{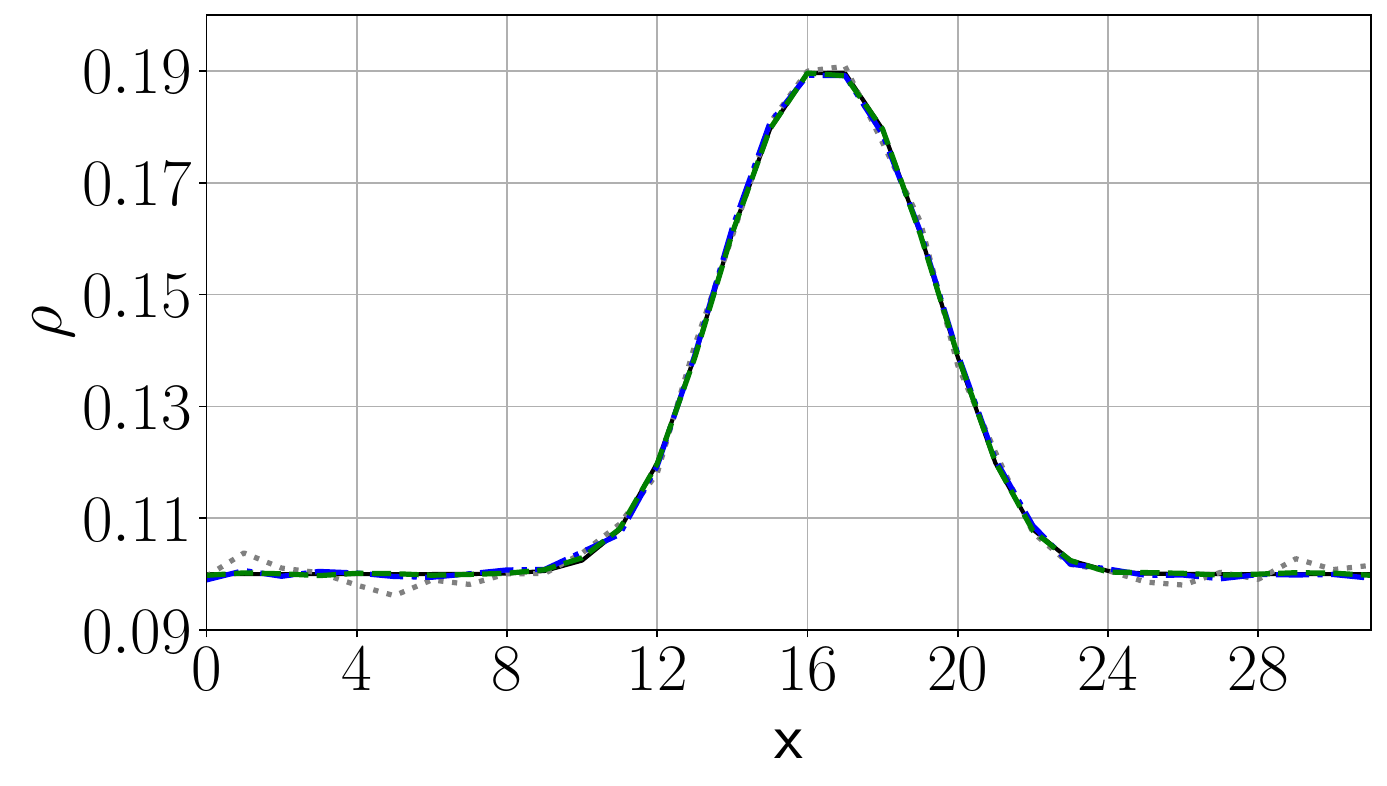}
        \caption{Result for 10 time steps.}
        \label{fig:D1Q3_10_time_steps}
    \end{subfigure}
    \hfill
    \begin{subfigure}{0.45\textwidth}
        \includegraphics[width=\textwidth]{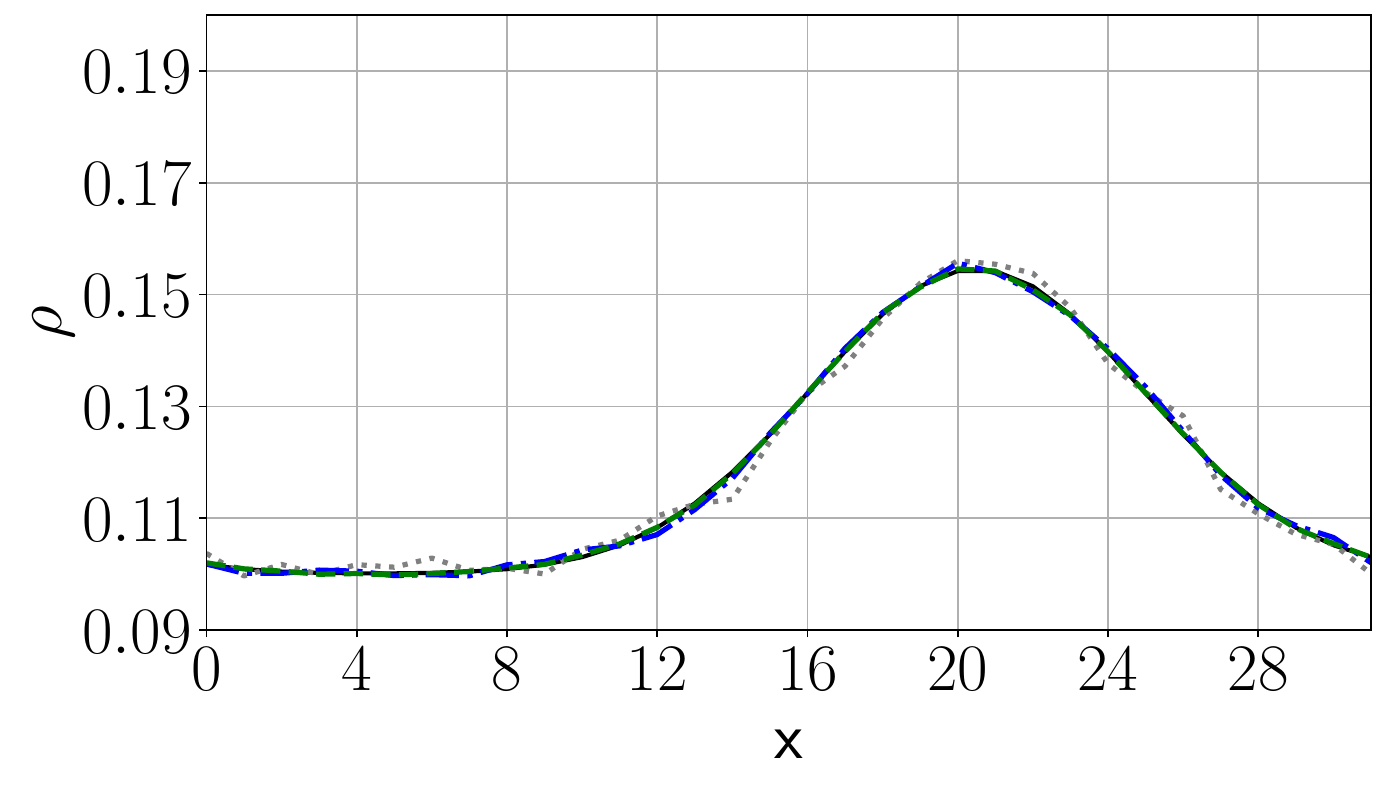}
        \caption{Result for 50 time steps.}
        \label{fig:D1Q3_50_time_steps}
    \end{subfigure}
    \\
    \begin{subfigure}{0.45\textwidth}
        \includegraphics[width=\textwidth]{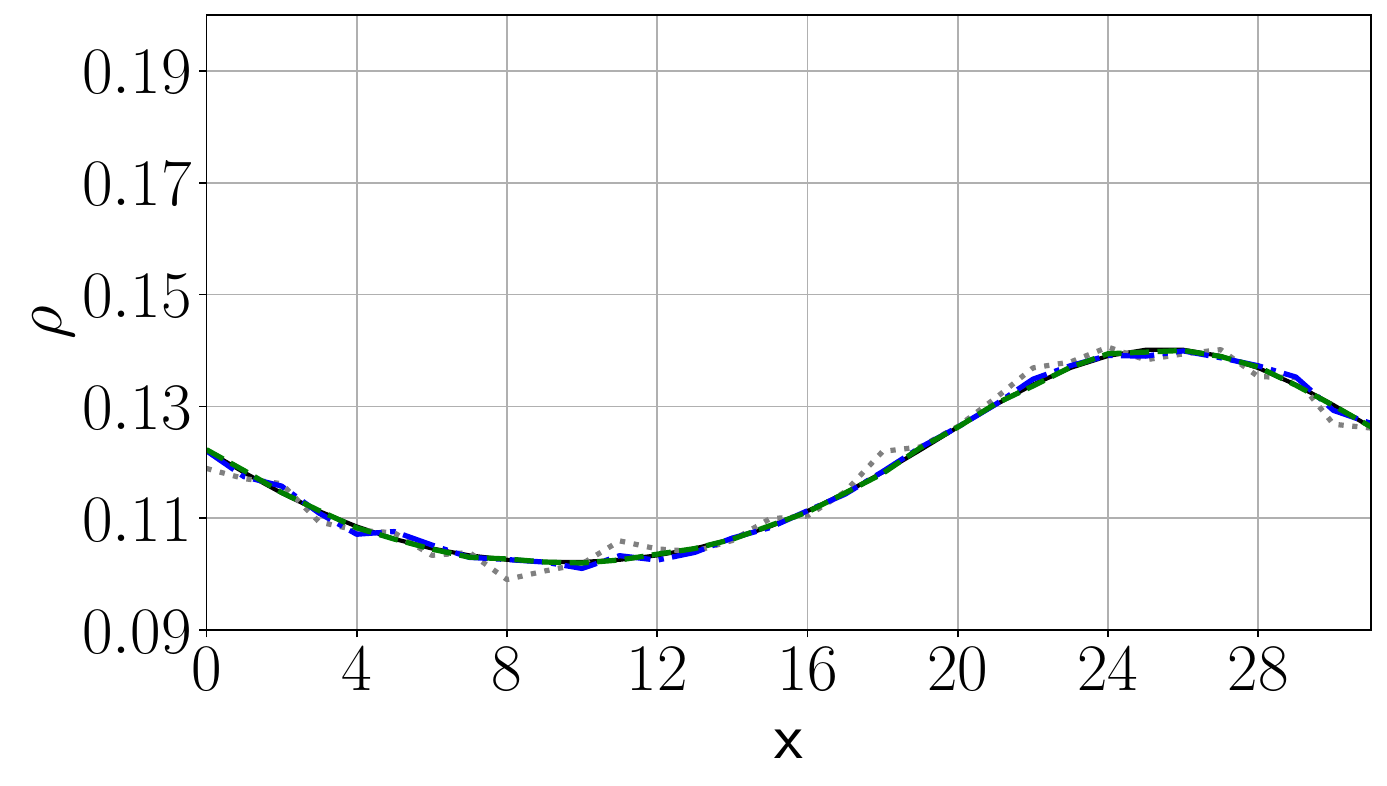}
        \caption{Result for 100 time steps.}
        \label{fig:D1Q3_100_time_steps}
    \end{subfigure}
    \hfill
    \begin{subfigure}{0.45\textwidth}
        \includegraphics[width=\textwidth]{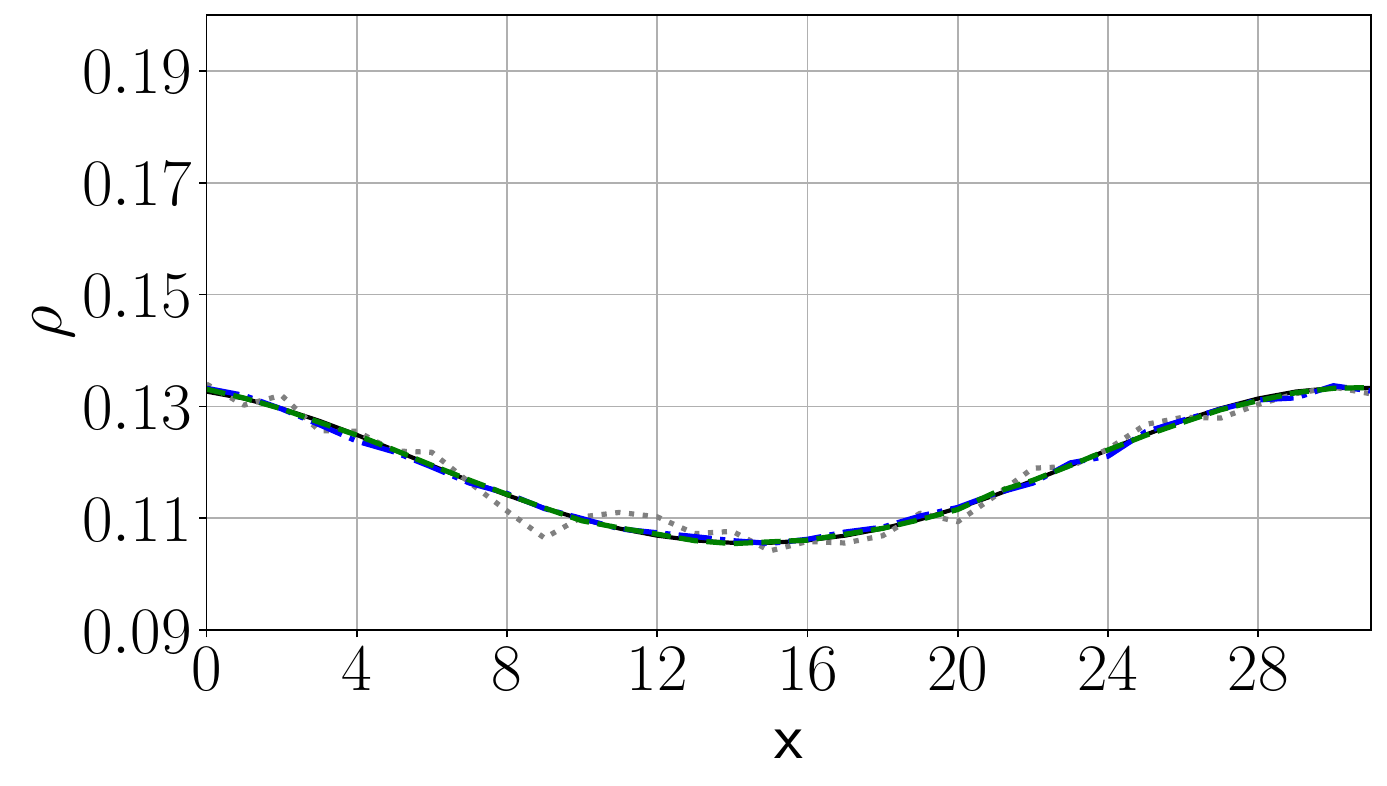}
        \caption{Result for 150 time steps.}
        \label{fig:D1Q3_150_time_steps}
    \end{subfigure}
    \\
    \begin{subfigure}{0.45\textwidth}
        \includegraphics[width=\textwidth]{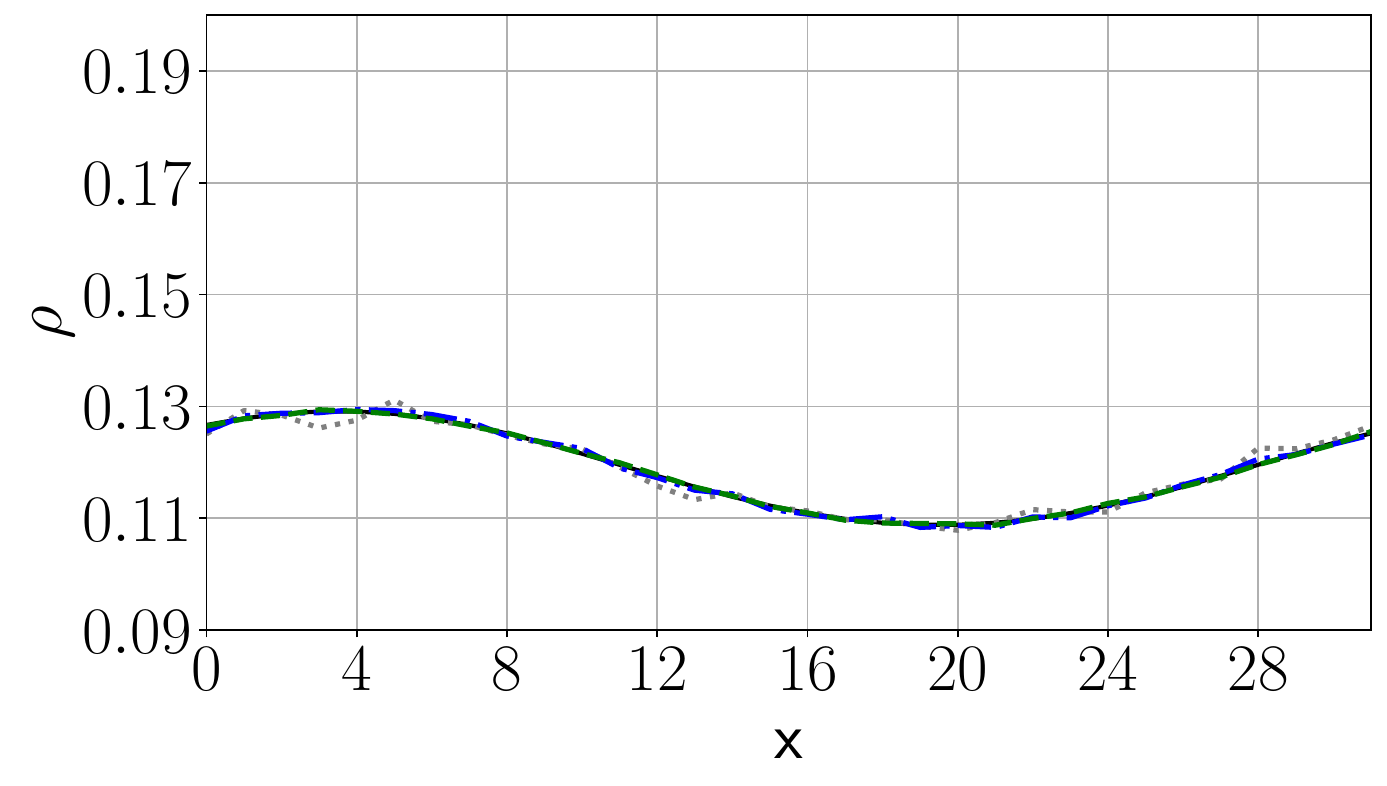}
        \caption{Result for 200 time steps.}
        \label{fig:D1Q3_200_time_steps}
    \end{subfigure}
    \hfill
    \begin{subfigure}{0.45\textwidth}
        \includegraphics[width=\textwidth]{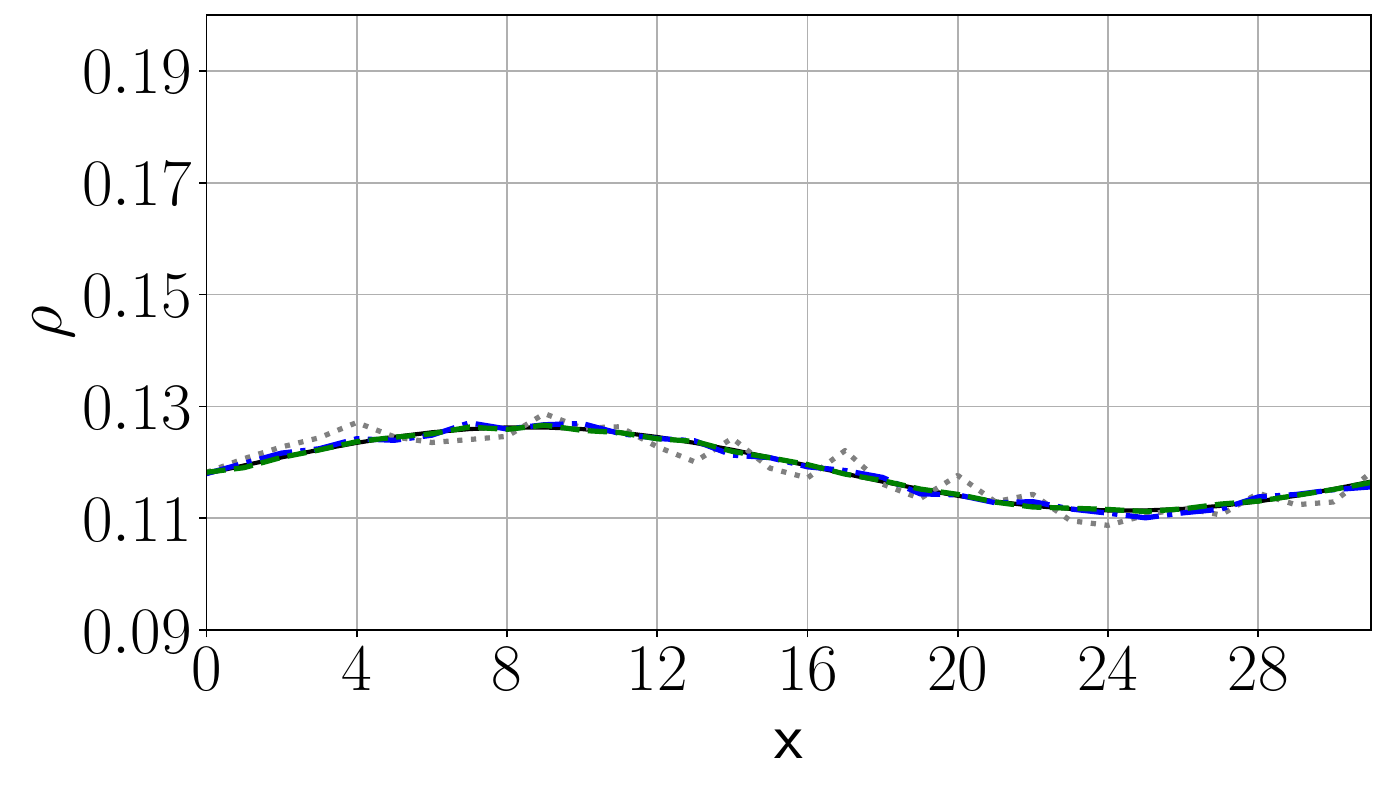}
        \caption{Result for 250 time steps.}
        \label{fig:D1Q3_250_time_steps}
    \end{subfigure}
    \caption{Results for QLBM and digital LBM using the D1Q3 velocity set for multiple time steps with different numbers of shots. 
 We compare the digital result (\fullblack{}), the QLBM result using $10^7$ shots (\dashedgreen{}), the QLBM result using $10^6$ shots (\dotdashblue{}), and the QLBM result using $10^5$ shots (\dottedgray{}).}
    \label{fig:D1Q3_multiple_t_var_shots}
\end{figure}

Based on the considerations outlined in \cref{sec.3.3}, the algorithmic complexity increases with the number of time steps. 
Consequently, a larger number of shots is required to maintain accuracy. This relationship is depicted in \cref{fig:D1Q3_multiple_t_var_shots}. 
While \cref{fig:D1Q3_10_time_steps} demonstrates that simulations with all tested shot counts closely align with the baseline solution for 10 time steps, the results for 250 time steps, shown in \cref{fig:D1Q3_250_time_steps}, reveal that $10^5$ and $10^6$ shots fail to accurately reproduce the digital baseline solution.
For simulations with an adequate number of shots, the algorithm reliably computes results over multiple time steps.

\begin{figure}
    \centering
    \includegraphics[width=\textwidth]{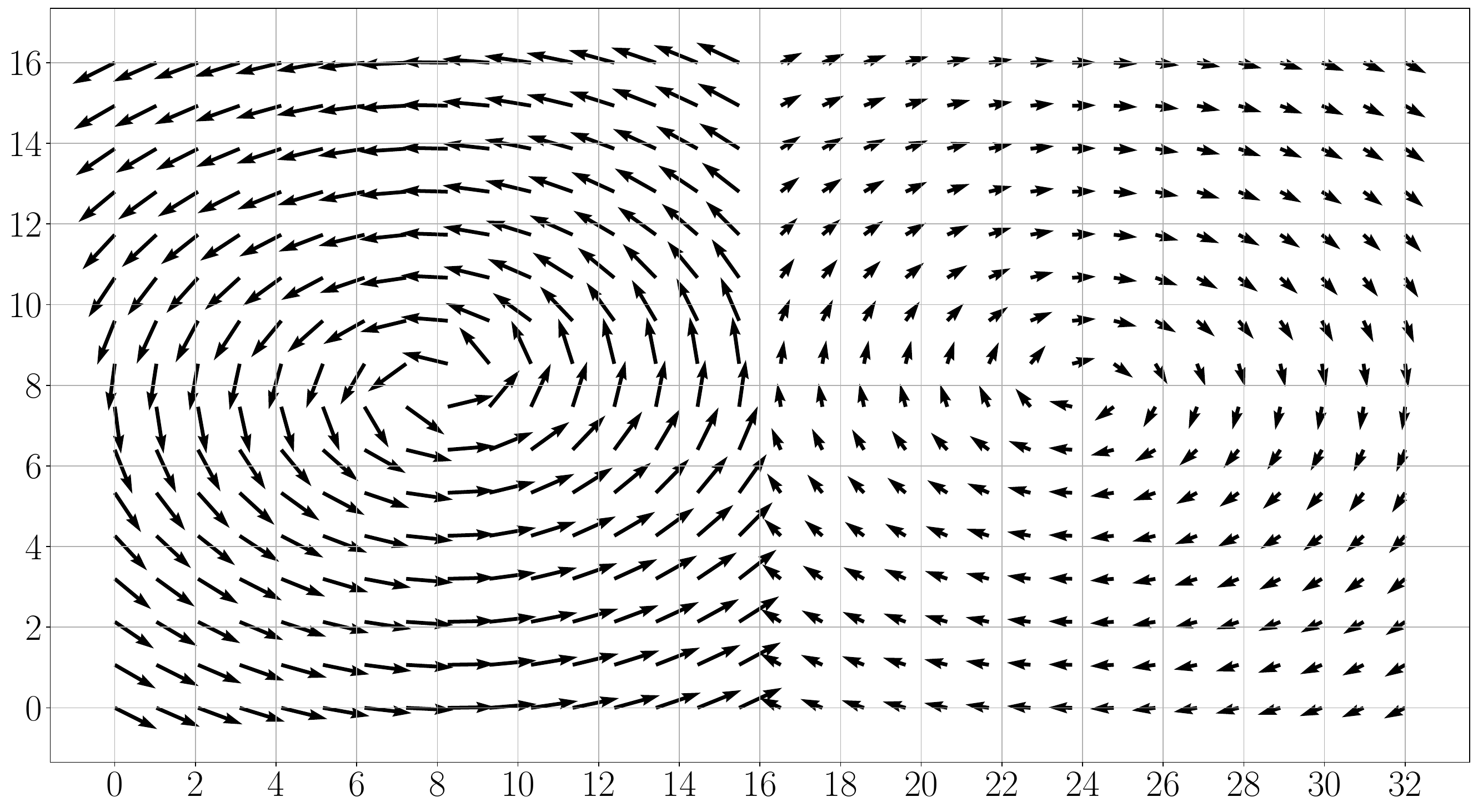}
    \caption{Advection velocity field used for the QLBM D2Q9 velocity set.}
    \label{fig:velocity_field_D2Q9}
\end{figure}

\begin{figure}
    \centering
    \begin{subfigure}{0.45\textwidth}
        \includegraphics[width=\textwidth]{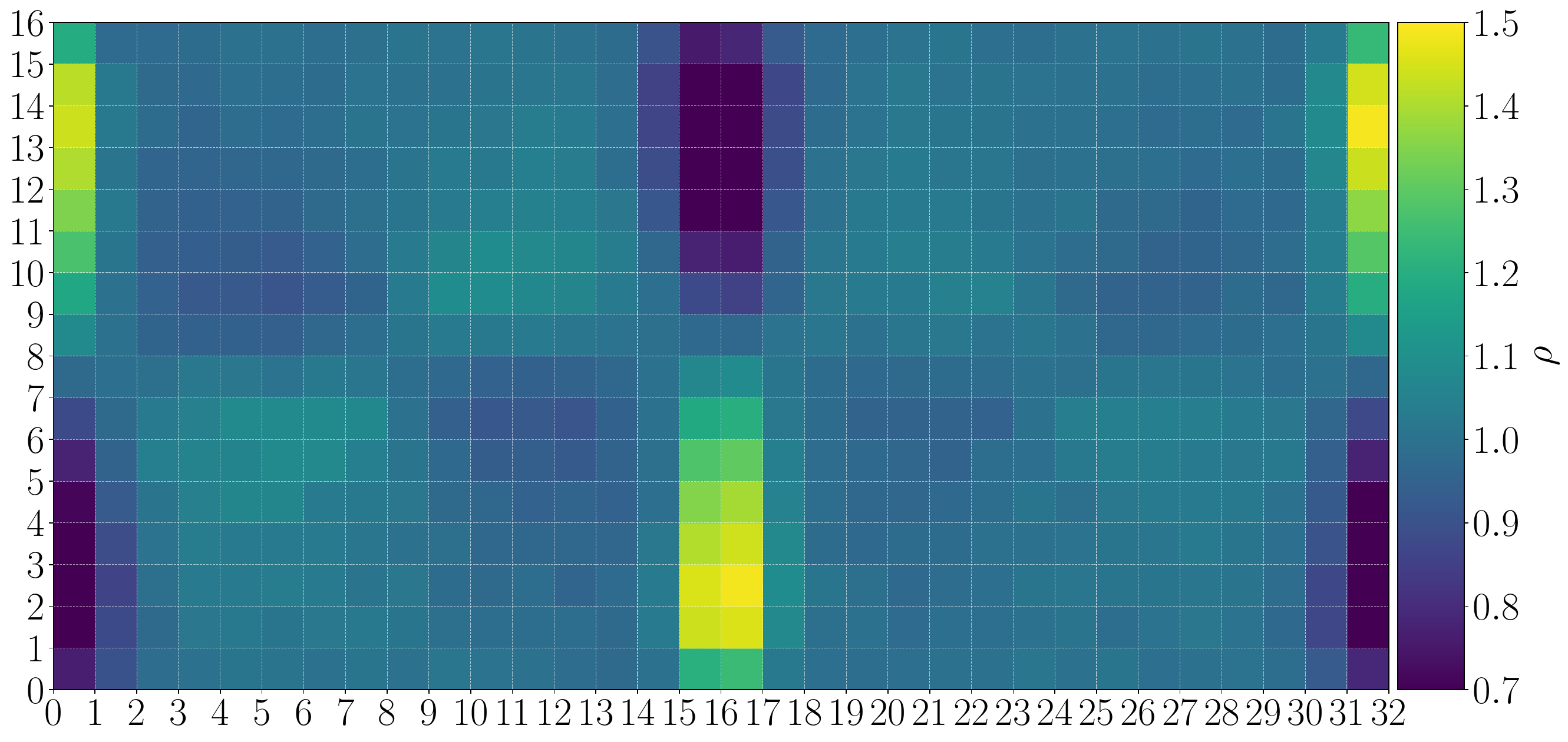}
        \caption{LBM result for 5 time steps.}
        \label{fig:D2Q9_5_time_steps_vortex_digital}
    \end{subfigure}
    \hfill
    \begin{subfigure}{0.45\textwidth}
        \includegraphics[width=\textwidth]{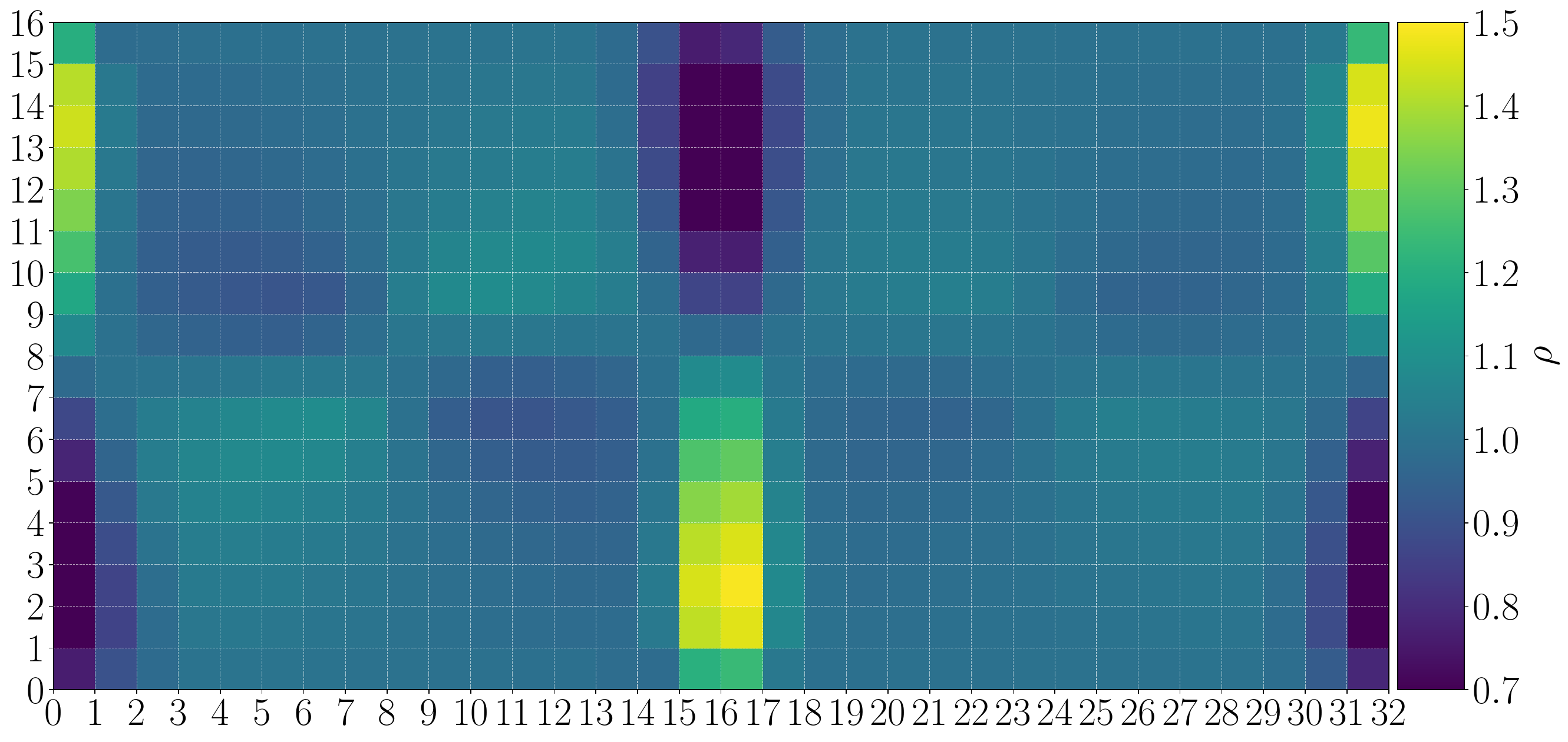}
        \caption{QLBM result for 5 time steps.}
        \label{fig:D2Q9_5_time_steps_vortex_quantum}
    \end{subfigure}
    \\
    \begin{subfigure}{0.45\textwidth}
        \includegraphics[width=\textwidth]{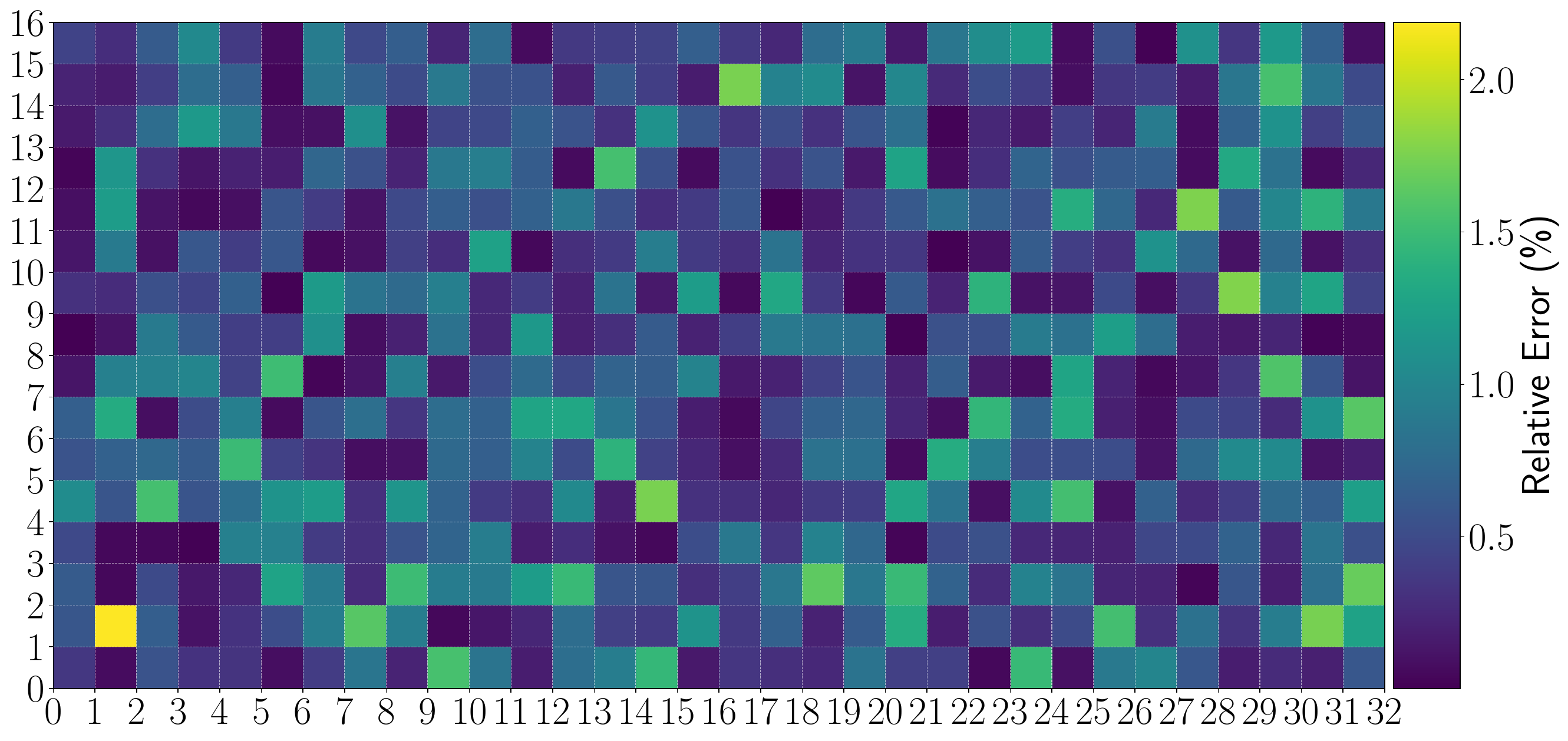}
        \caption{Relative error between LBM and QLBM.}
        \label{fig:D2Q9_5_time_steps_vortex_rel_error}
    \end{subfigure}
    \hfill
    \begin{subfigure}{0.45\textwidth}
        \includegraphics[width=\textwidth]{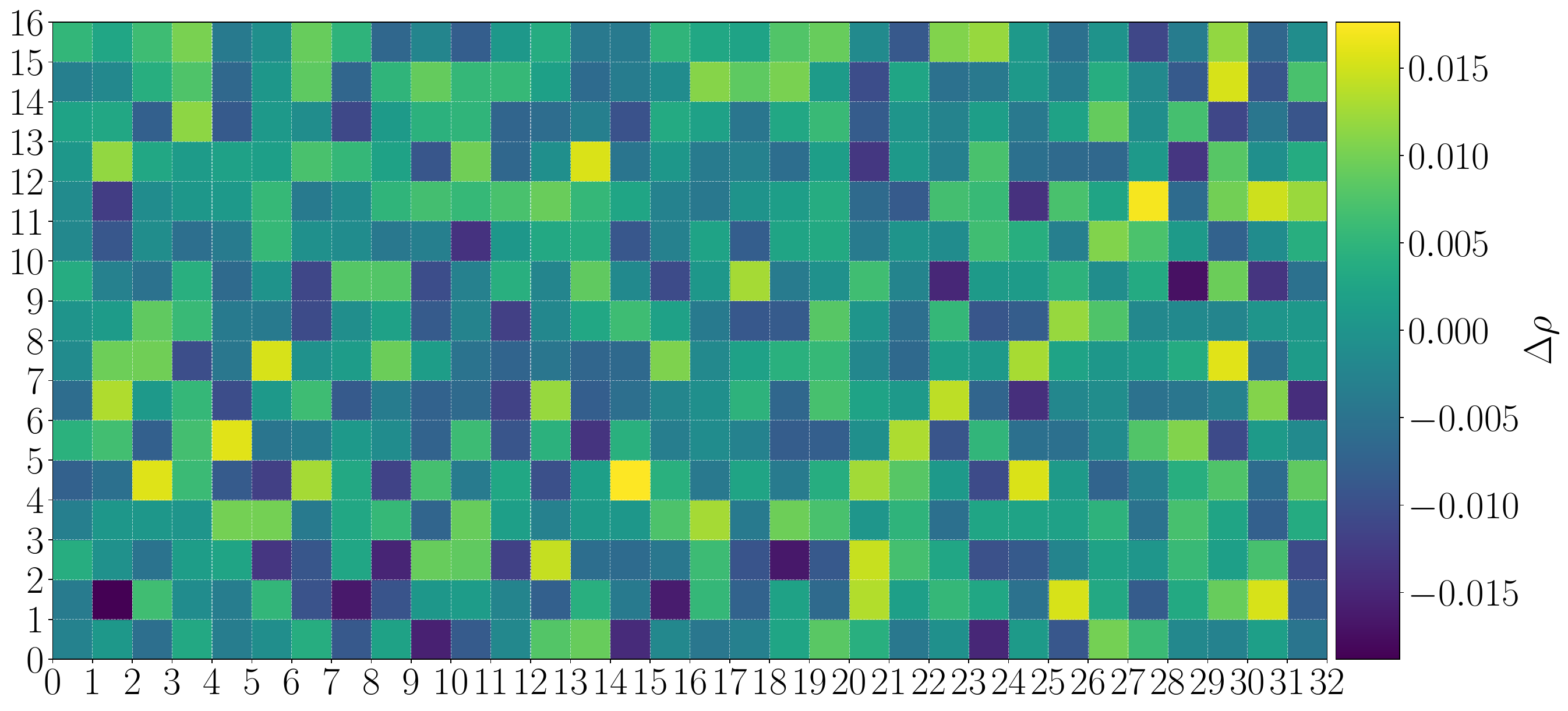}
        \caption{Difference between LBM and QLBM.}
        \label{fig:D2Q9_5_time_steps_vortex_diff}
    \end{subfigure}
    \caption{Results and error for QLBM and digital LBM using the D2Q9 velocity set for 5 time steps with $10^7$ shots and a double vortex advection velocity.
 The MAPE for 5 time steps is 0.573\%.
 }
    \label{fig:D2Q9_t_5_vortex}
\end{figure}

\begin{figure}
    \centering
    \begin{subfigure}{0.45\textwidth}
        \includegraphics[width=\textwidth]{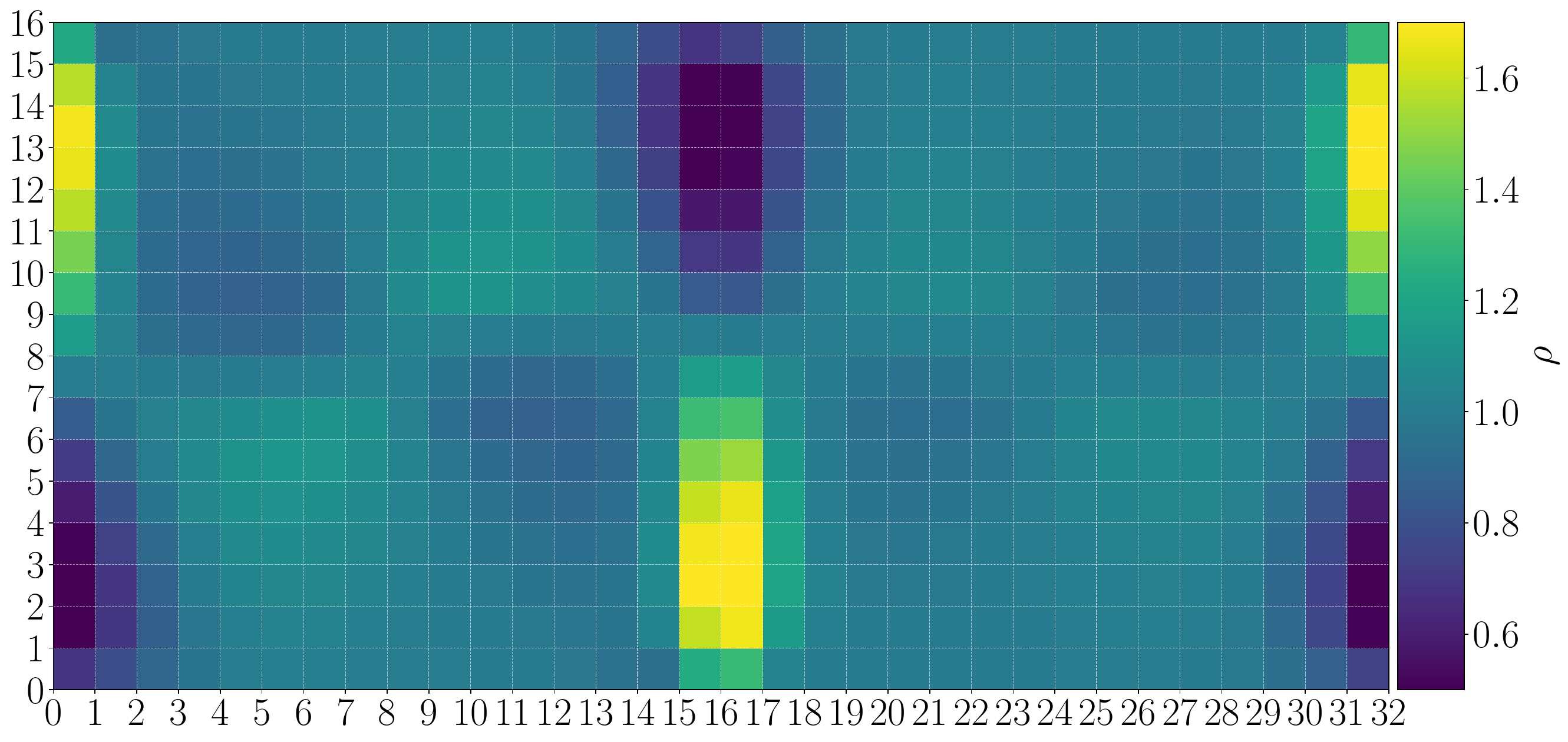}
        \caption{LBM result for 10 time steps.}
        \label{fig:D2Q9_10_time_steps_vortex_digital}
    \end{subfigure}
    \hfill
    \begin{subfigure}{0.45\textwidth}
        \includegraphics[width=\textwidth]{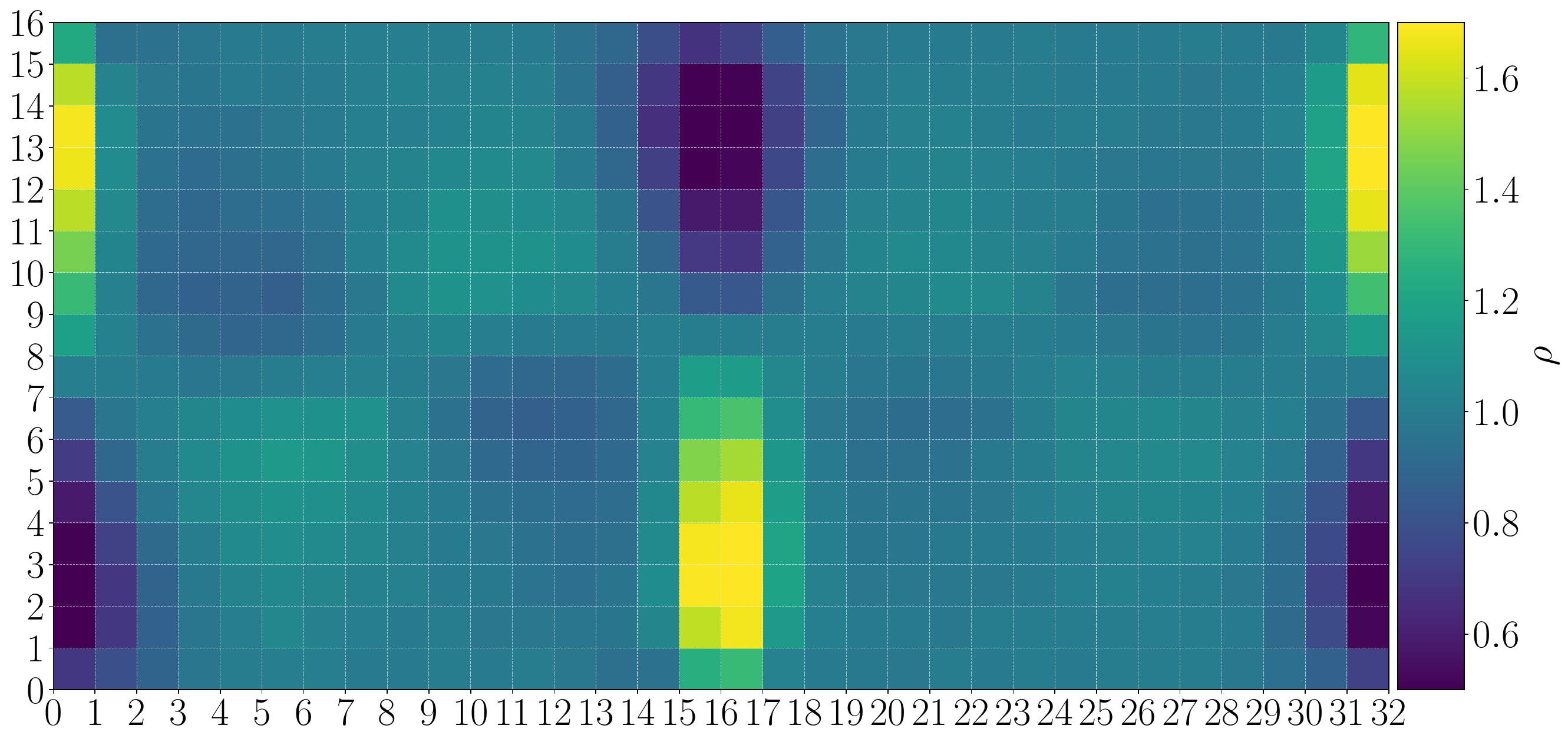}
        \caption{QLBM result for 10 time steps.}
        \label{fig:D2Q9_10_time_steps_vortex_quantum}
    \end{subfigure}
    \\
    \begin{subfigure}{0.45\textwidth}
        \includegraphics[width=\textwidth]{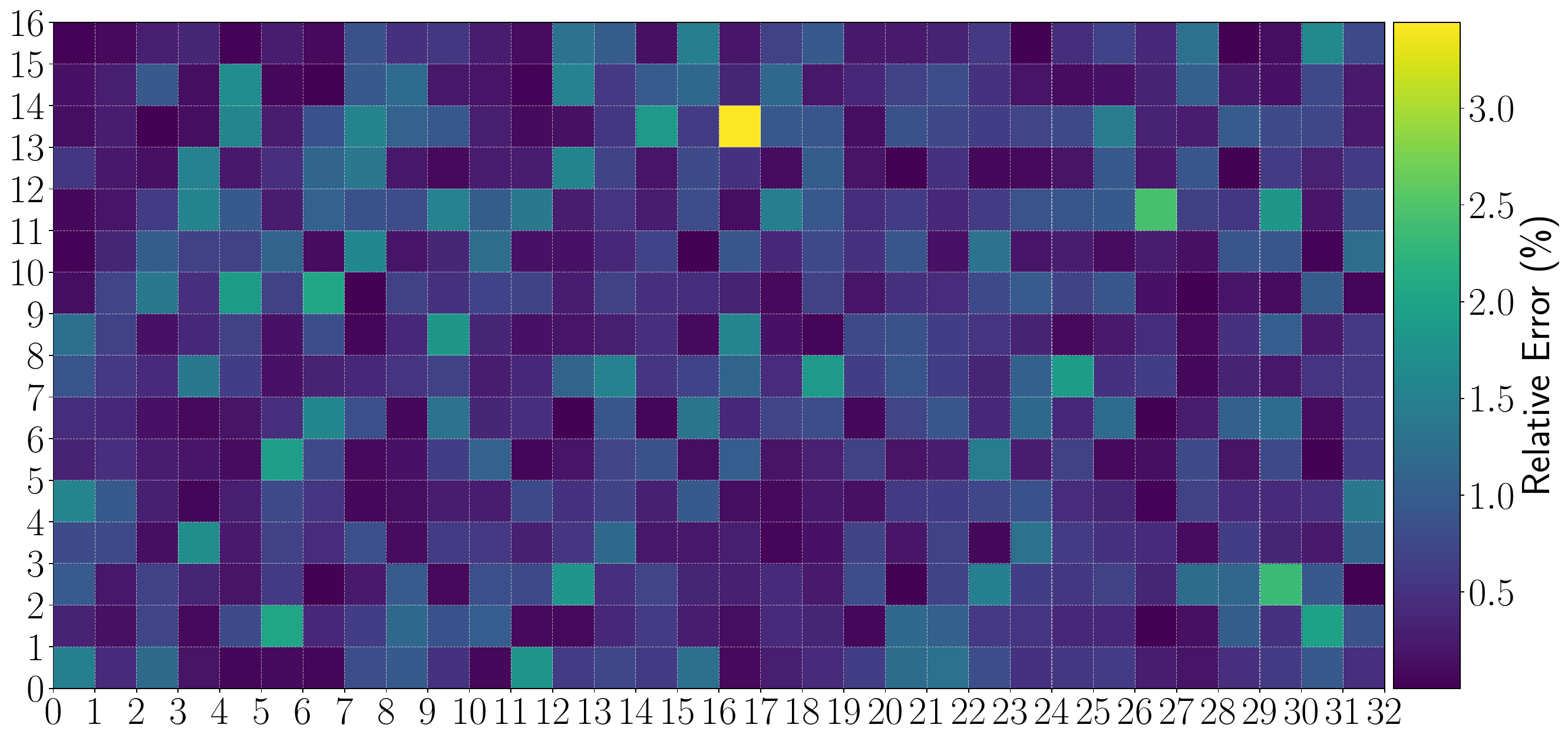}
        \caption{Relative error between LBM and QLBM.}
        \label{fig:D2Q9_10_time_steps_vortex_rel_error}
    \end{subfigure}
    \hfill
    \begin{subfigure}{0.45\textwidth}
        \includegraphics[width=\textwidth]{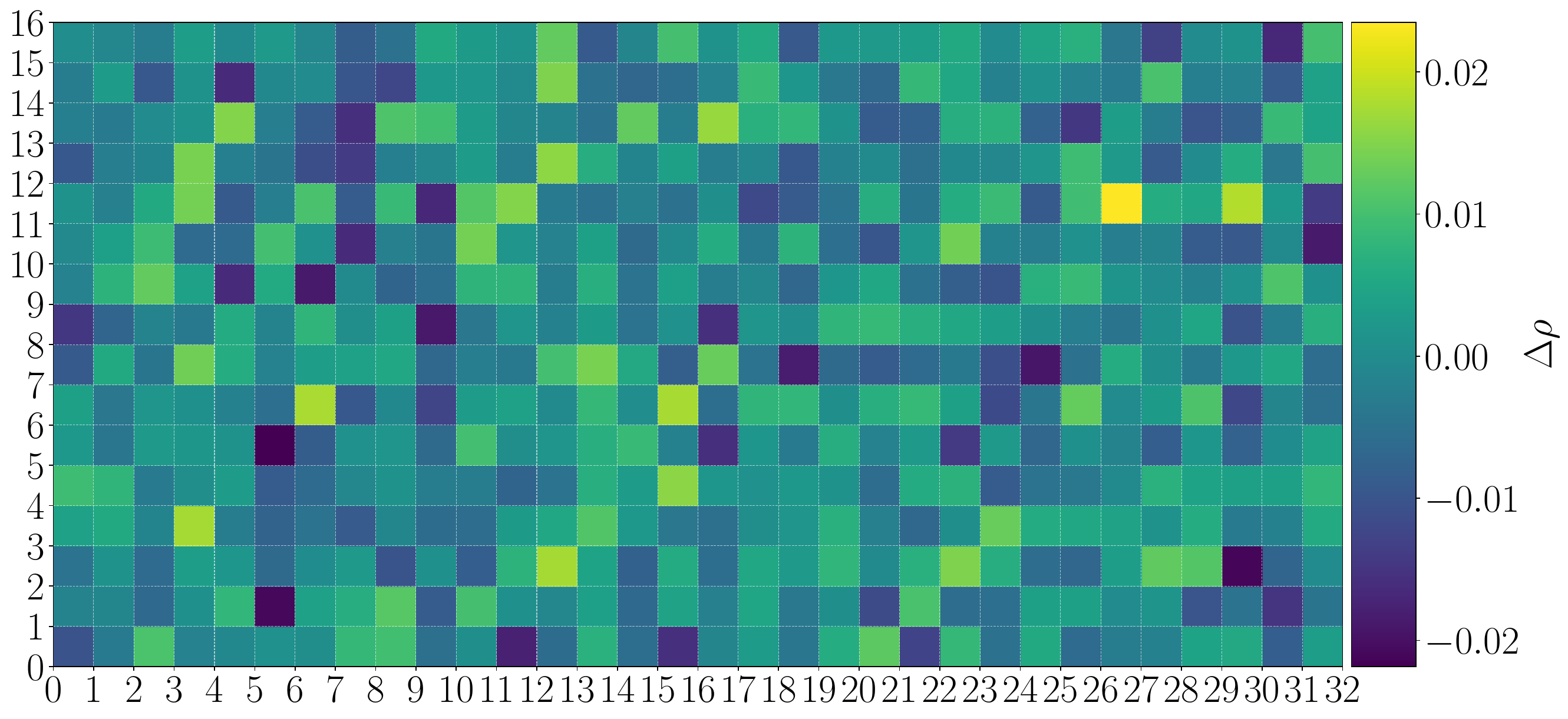}
        \caption{Difference between LBM and QLBM.}
        \label{fig:D2Q9_10_time_steps_vortex_diff}
    \end{subfigure}
    \caption{Results and error for QLBM and digital LBM using the D2Q9 velocity set for 10 time steps with $10^7$ shots and a double vortex advection velocity.
 The MAPE for 10 time steps is 0.593\%.
 }
    \label{fig:D2Q9_t_10_vortex}
\end{figure}

\begin{figure}
    \centering
    \begin{subfigure}{0.45\textwidth}
        \includegraphics[width=\textwidth]{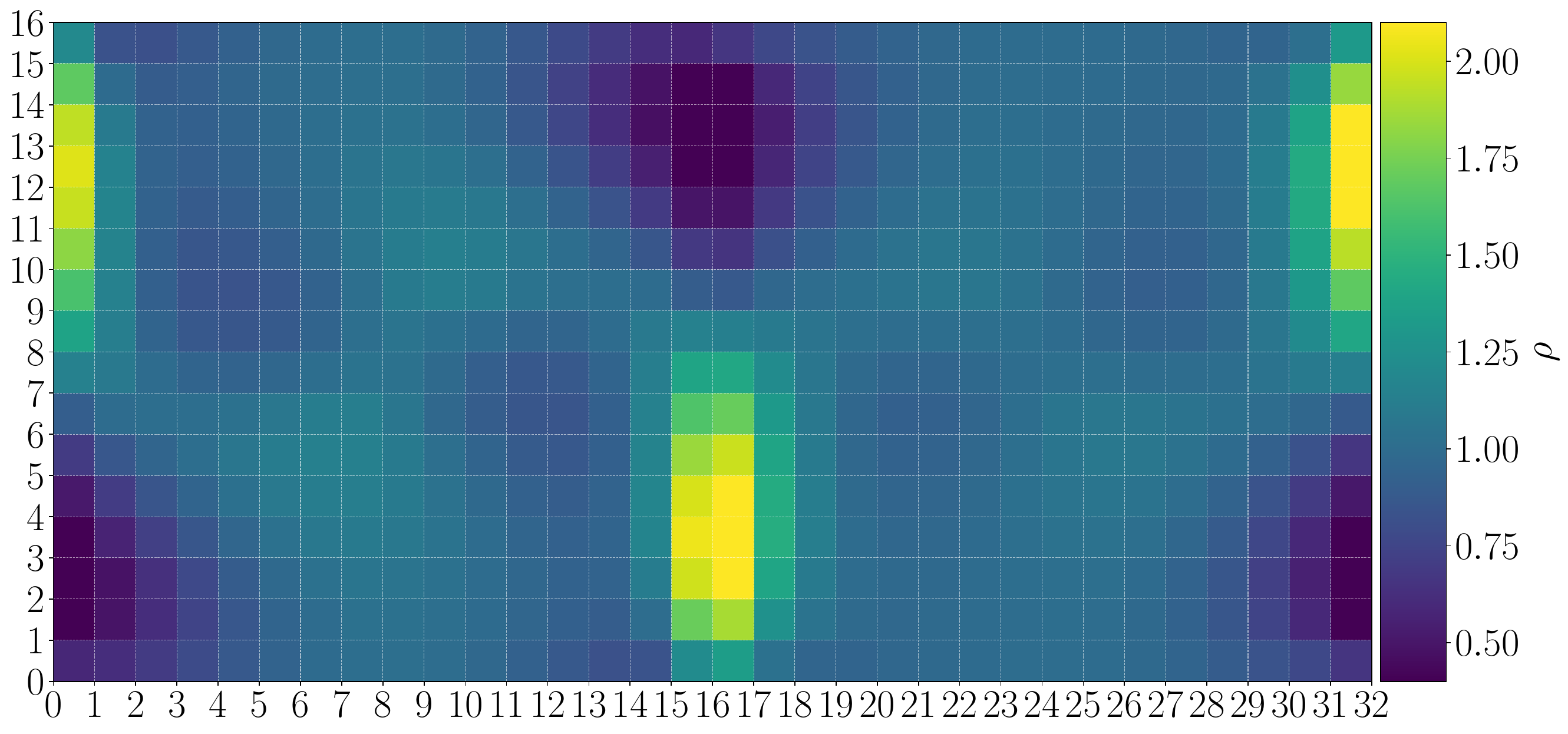}
        \caption{LBM result for 25 time steps.}
        \label{fig:D2Q9_25_time_steps_vortex_digital}
    \end{subfigure}
    \hfill
    \begin{subfigure}{0.45\textwidth}
        \includegraphics[width=\textwidth]{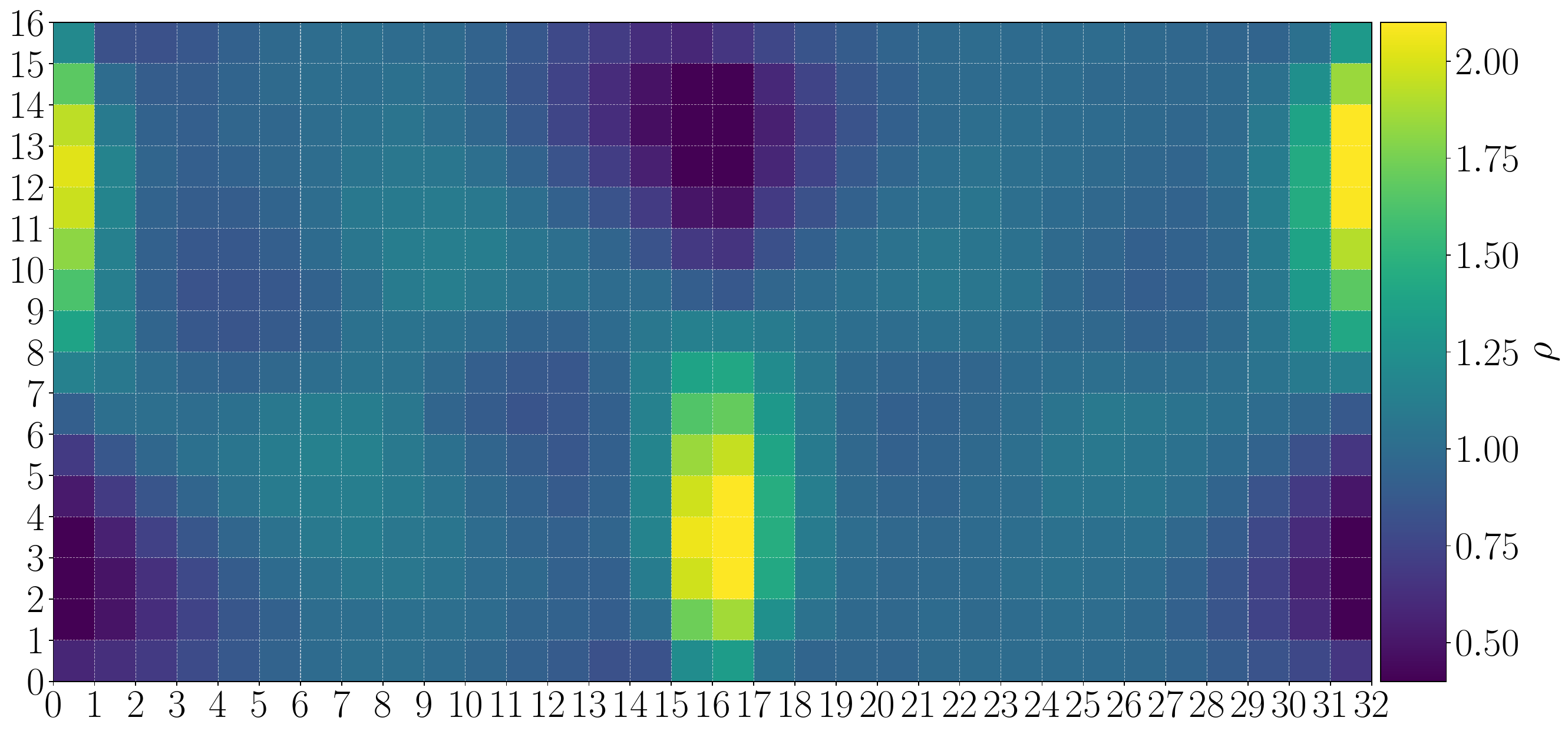}
        \caption{QLBM result for 25 time steps.}
        \label{fig:D2Q9_25_time_steps_vortex_quantum}
    \end{subfigure}
    \\
    \begin{subfigure}{0.45\textwidth}
        \includegraphics[width=\textwidth]{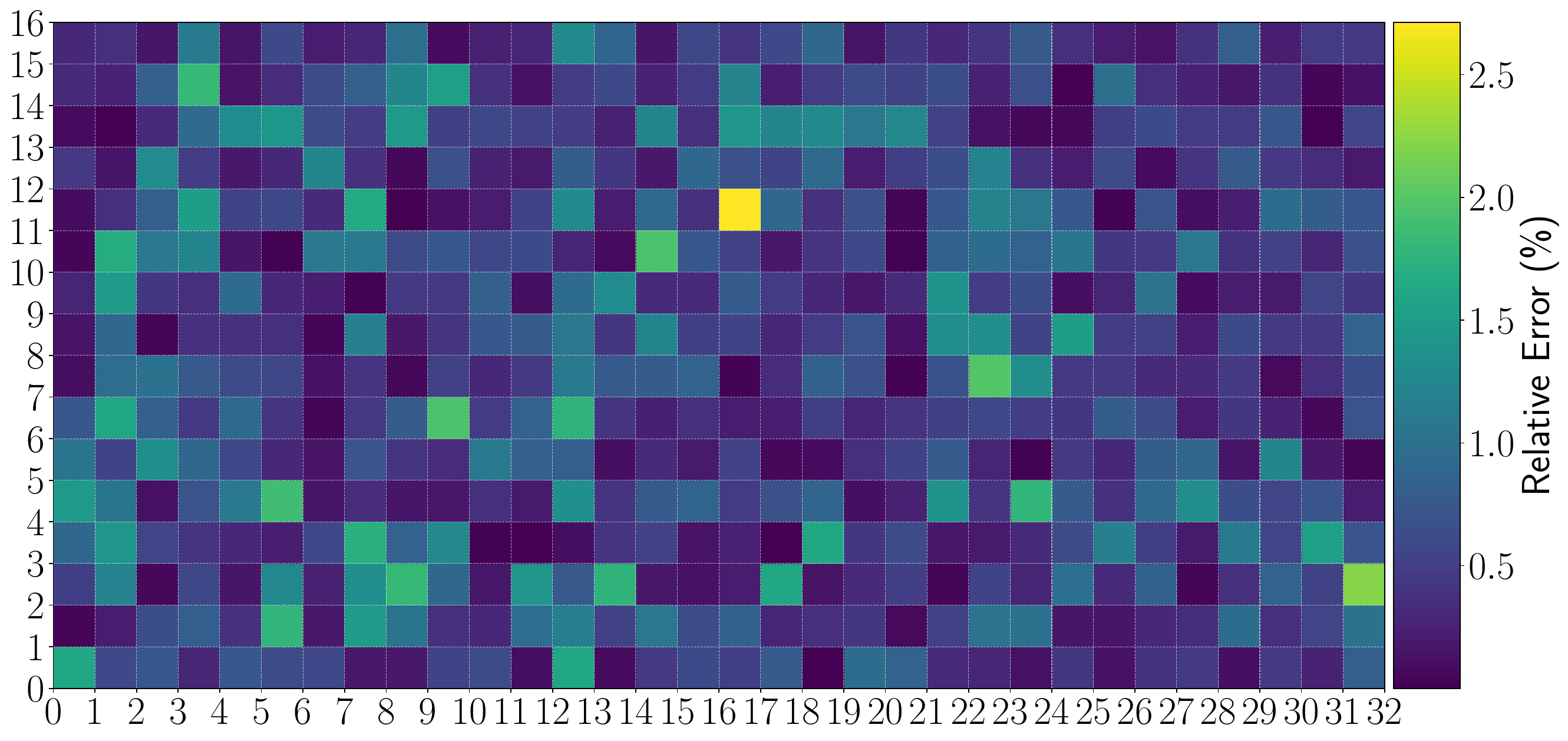}
        \caption{Relative error between LBM and QLBM.}
        \label{fig:D2Q9_25_time_steps_vortex_rel_error}
    \end{subfigure}
    \hfill
    \begin{subfigure}{0.45\textwidth}
        \includegraphics[width=\textwidth]{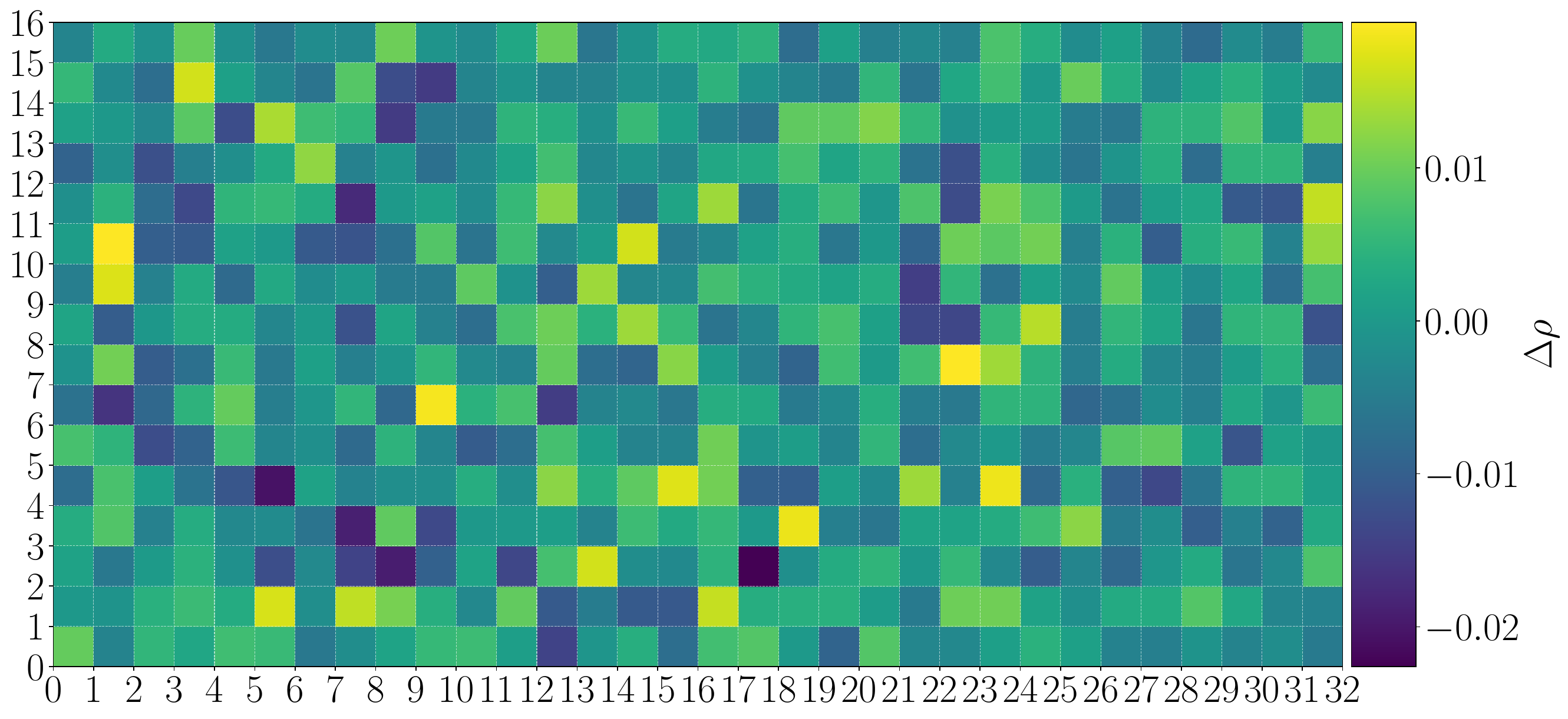}
        \caption{Difference between LBM and QLBM.}
        \label{fig:D2Q9_25_time_steps_vortex_diff}
    \end{subfigure}
    \caption{Results and error for QLBM and digital LBM using the D2Q9 velocity set for 25 time steps with $10^7$ shots and a double vortex advection velocity.
 The MAPE for 25 time steps is 0.594\%.
 }
    \label{fig:D2Q9_t_25_vortex}
\end{figure}

For the two-dimensional case, a grid resolution of $[32 \times 16]$ points is employed. 
The density is initialized uniformly as $\rho(x, y) = 1$. 
The initialization of the quantum circuit is carried out by generating a uniform probability distribution, implemented via the application of a Hadamard gate to each qubit. 
The advection velocity field, representing a double vortex configuration, is defined as 
\begin{equation}
    \begin{aligned}
 u(x, y) &=
    \begin{cases}
 - S_1 \frac{y - y_1}{\sqrt{(x - x_1)^2 + (y - y_1)^2 + \epsilon}}, & x \leq \frac{L_x}{2}, \\
 S_2 \frac{y - y_2}{\sqrt{(x - x_2)^2 + (y - y_2)^2 + \epsilon}}, & x > \frac{L_x}{2},
    \end{cases} \\
 v(x, y) &=
    \begin{cases}
 S_1 \frac{x - x_1}{\sqrt{(x - x_1)^2 + (y - y_1)^2 + \epsilon}}, & x \leq \frac{L_x}{2}, \\
 - S_2 \frac{x - x_2}{\sqrt{(x - x_2)^2 + (y - y_2)^2 + \epsilon}}, & x > \frac{L_x}{2},
    \end{cases}
    \end{aligned}
\end{equation}
where $u$ and $v$ denote the velocity components in the $x$- and $y$-directions, respectively. 
The parameters are specified as follows: $S_1=0.2$ and $S_2=0.1$ represent the strengths of the left and right vortices, $(x_1, y_1) = (0.25, 0.5)$ and $(x_2, y_2) = (0.75, 0.5)$ denote the normalized coordinates of the vortex centers, $\epsilon = 10^{-8}$ is a regularization parameter, and $L_x$ and $L_y$ are the domain lengths in the $x$- and $y$-directions, respectively. 
The resulting velocity field is illustrated in \cref{fig:velocity_field_D2Q9}.
Simulations are performed for $t = [5, 10, 25]$ time steps using $10^7$ shots. 
The simulation results and associated errors for the LBM and QLBM are presented in \cref{fig:D2Q9_t_5_vortex,fig:D2Q9_t_10_vortex,fig:D2Q9_t_25_vortex}. 
The QLBM results demonstrate a good agreement with the reference data. 
The observed discrepancies are attributed to random shot noise, which decreases with an increasing number of shots, as shown in \cref{fig:MAPEoneTimestep}.


As a third validation case, we consider a one-dimensional domain with a uniform initial density, $\rho(x, t=0) = 0.1$. 
The initialization of the quantum circuit is achieved by generating a uniform probability distribution, which is implemented by applying a Hadamard gate on each qubit.

We impose a linearly increasing advection velocity field given by $u(x) = 0.1x + 0.1$, where $x \in [0,1]$ represents normalized spatial coordinates. 
Multiple time steps are computed using the QLBM with a fixed number of shots set at $10^7$.

\begin{figure}
    \centering
    \begin{subfigure}{0.4\textwidth}
        \includegraphics[width=\textwidth]{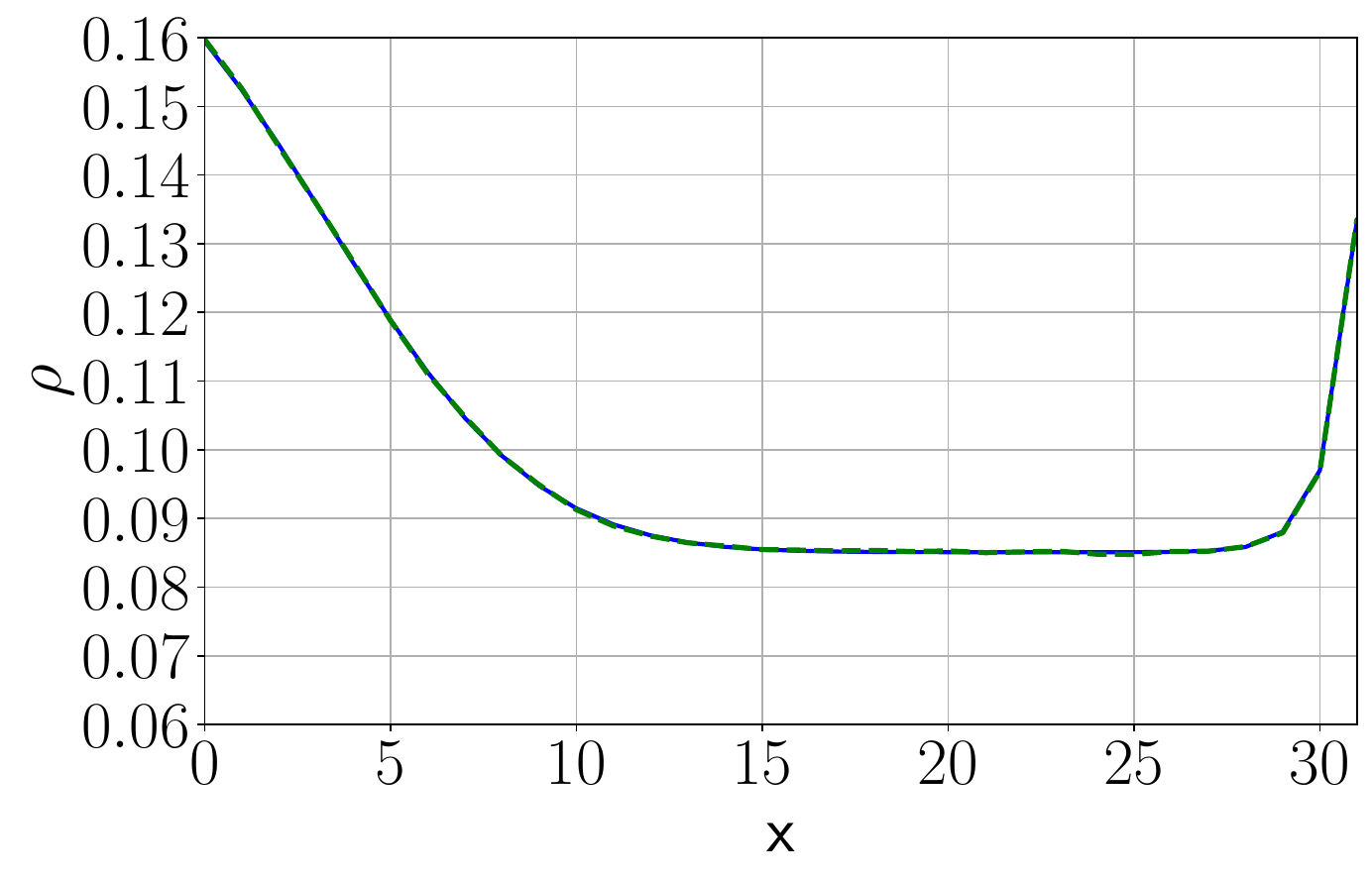}
        \caption{Result for 50 time steps.}
        \label{fig:D1Q3_50_time_steps_non_uniform_u}
    \end{subfigure}
    \hfill
    \begin{subfigure}{0.4\textwidth}
        \includegraphics[width=\textwidth]{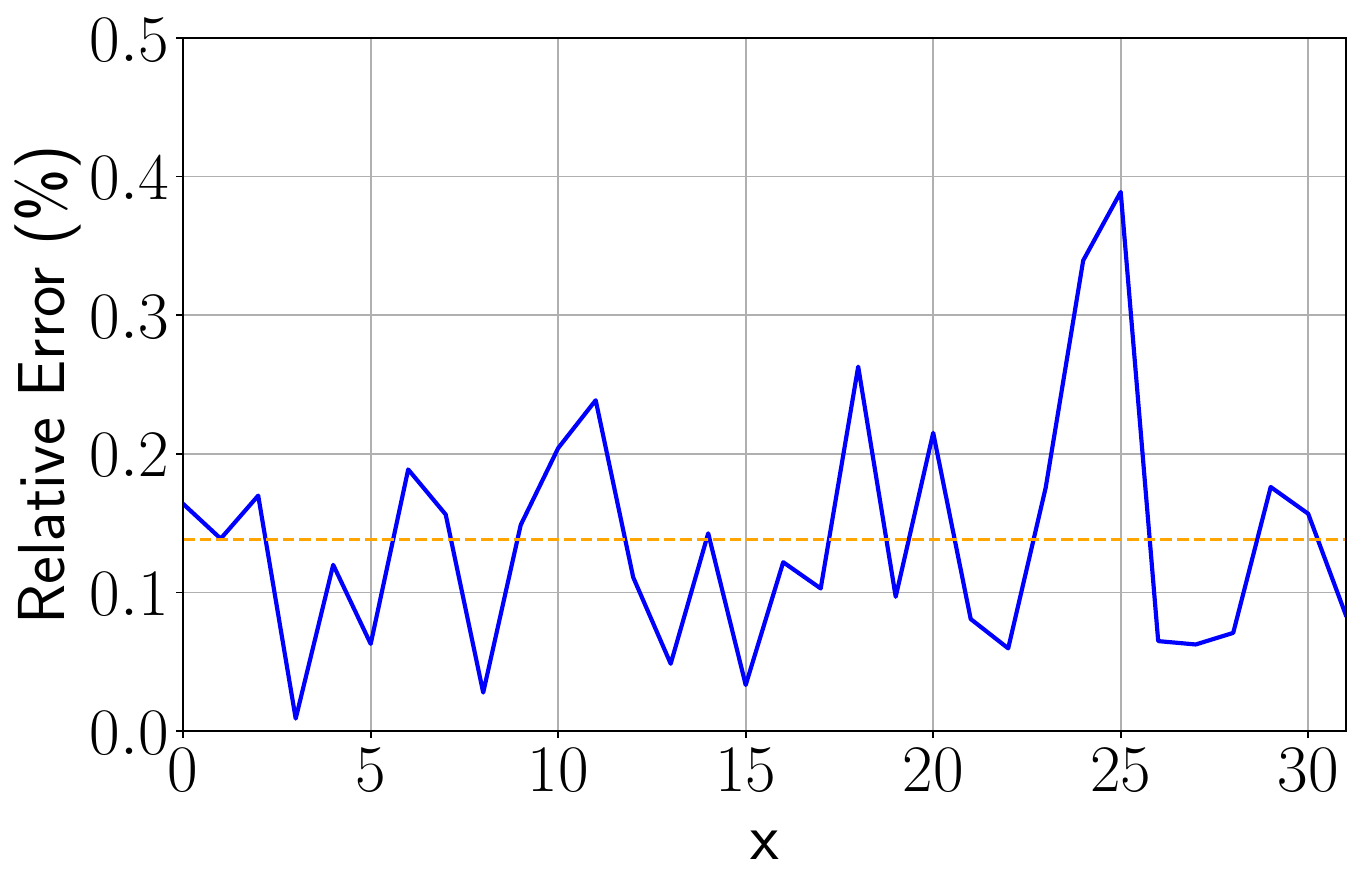}
        \caption{Relative error for 50 time steps.}
        \label{fig:D1Q3_50_time_steps_non_uniform_u_error}
    \end{subfigure}
    \\
    \begin{subfigure}{0.4\textwidth}
        \includegraphics[width=\textwidth]{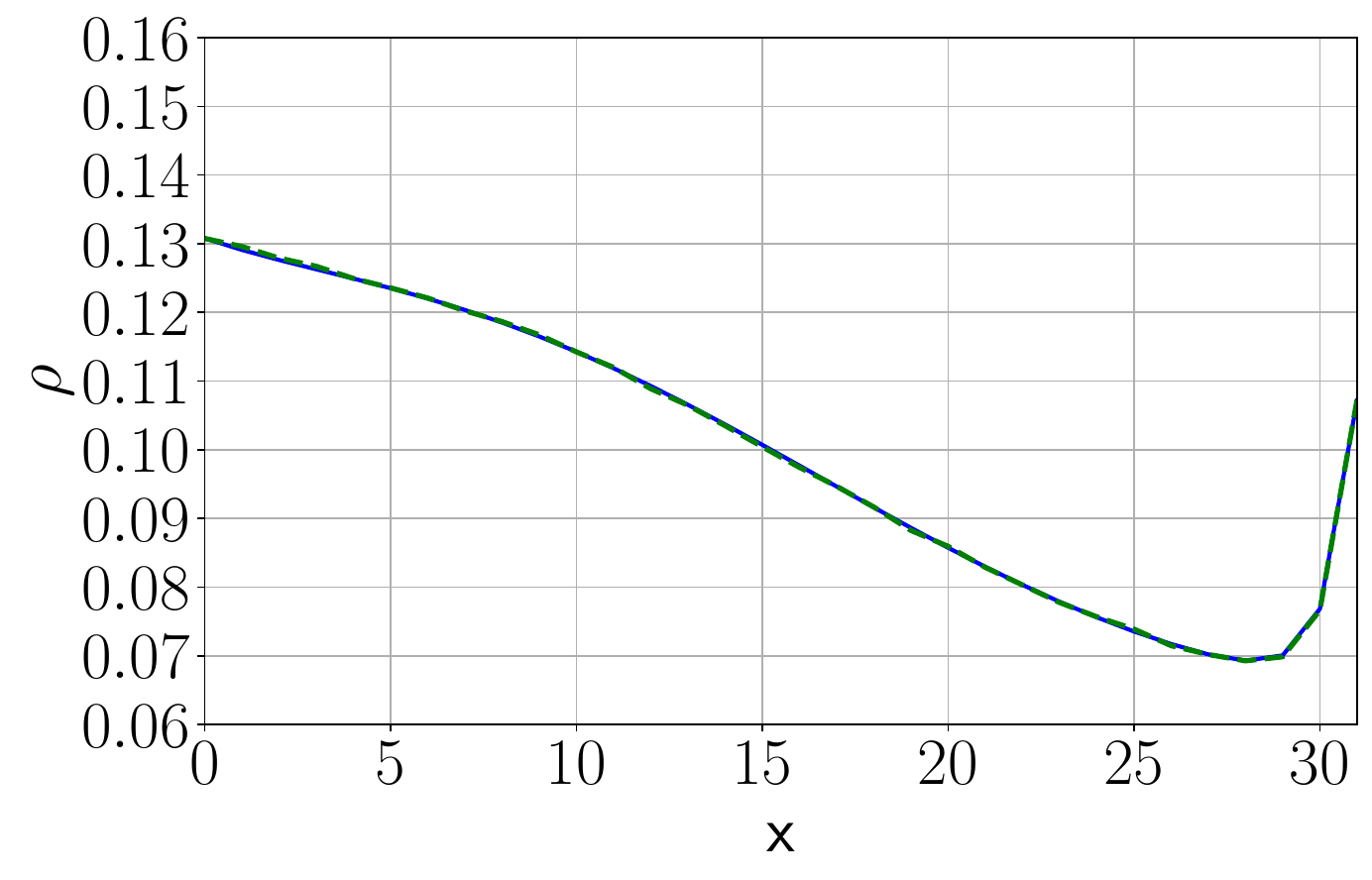}
        \caption{Result for 150 time steps.}
        \label{fig:D1Q3_150_time_steps_non_uniform_u}
    \end{subfigure}
    \hfill
    \begin{subfigure}{0.4\textwidth}
        \includegraphics[width=\textwidth]{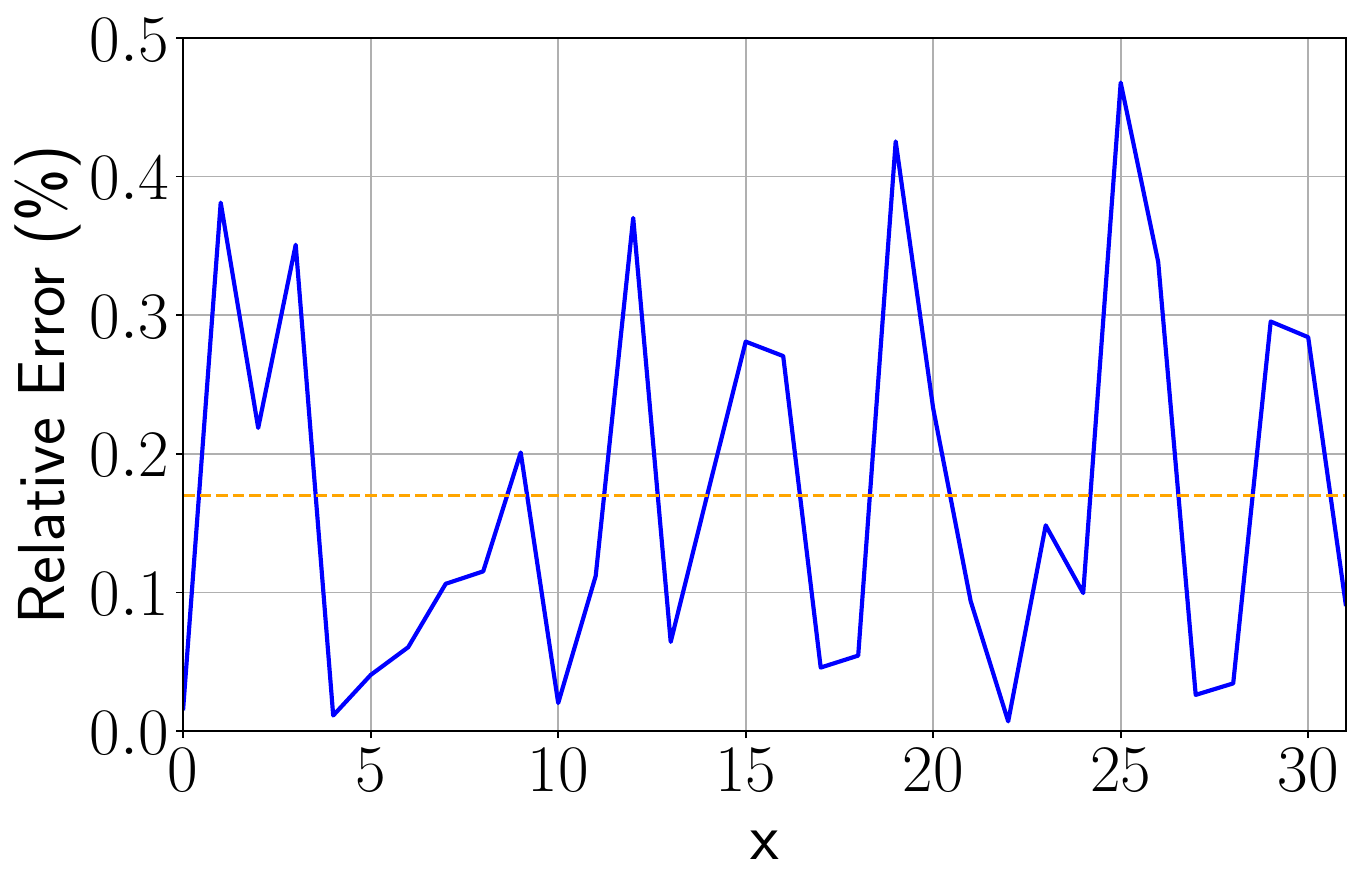}
        \caption{Relative error for 150 time steps.}
        \label{fig:D1Q3_150_time_steps_non_uniform_u_error}
    \end{subfigure}
    \\
    \begin{subfigure}{0.4\textwidth}
        \includegraphics[width=\textwidth]{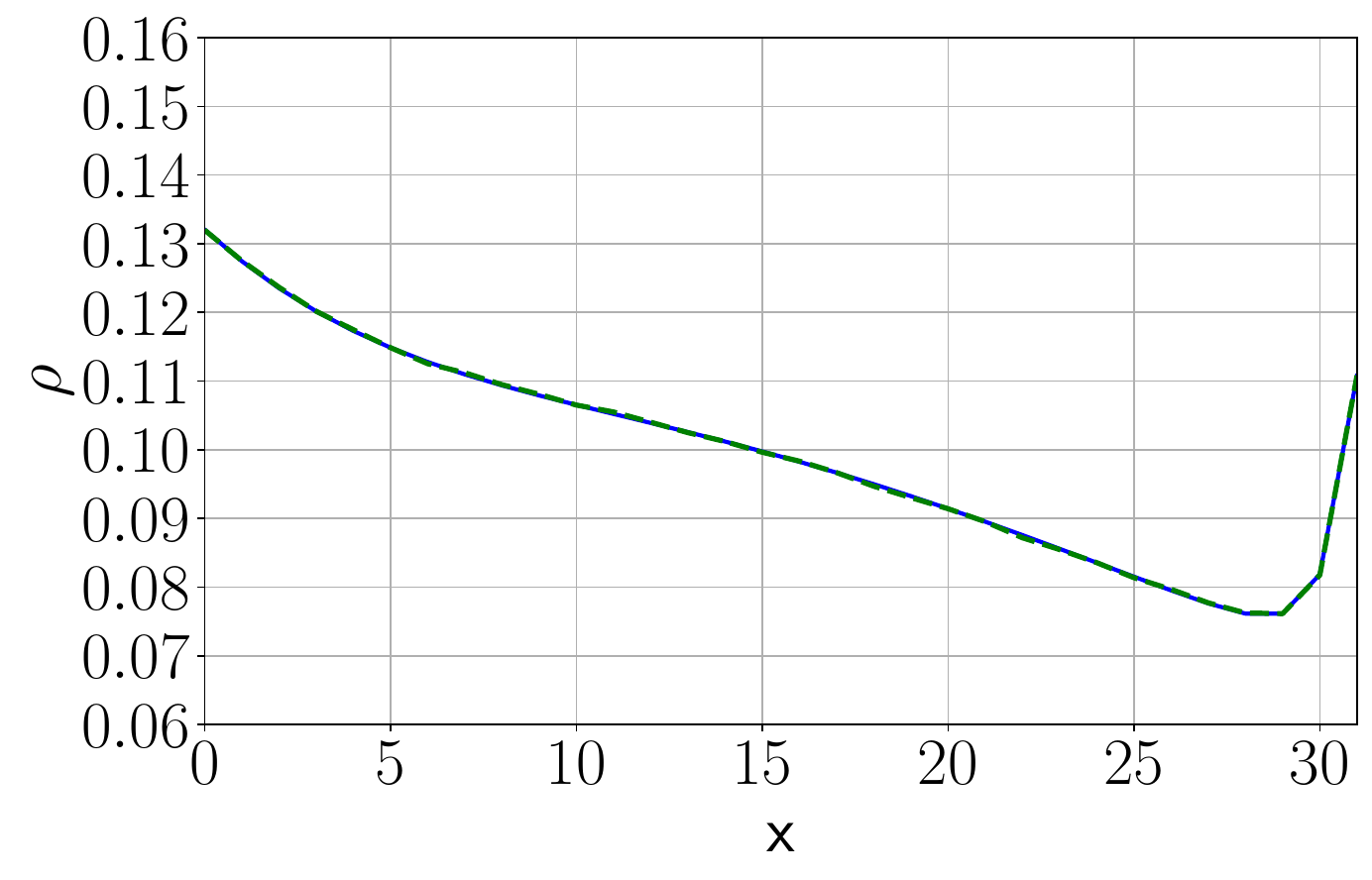}
        \caption{Result for 200 time steps.}
        \label{fig:D1Q3_200_time_steps_non_uniform_u}
    \end{subfigure}
    \hfill
    \begin{subfigure}{0.4\textwidth}
        \includegraphics[width=\textwidth]{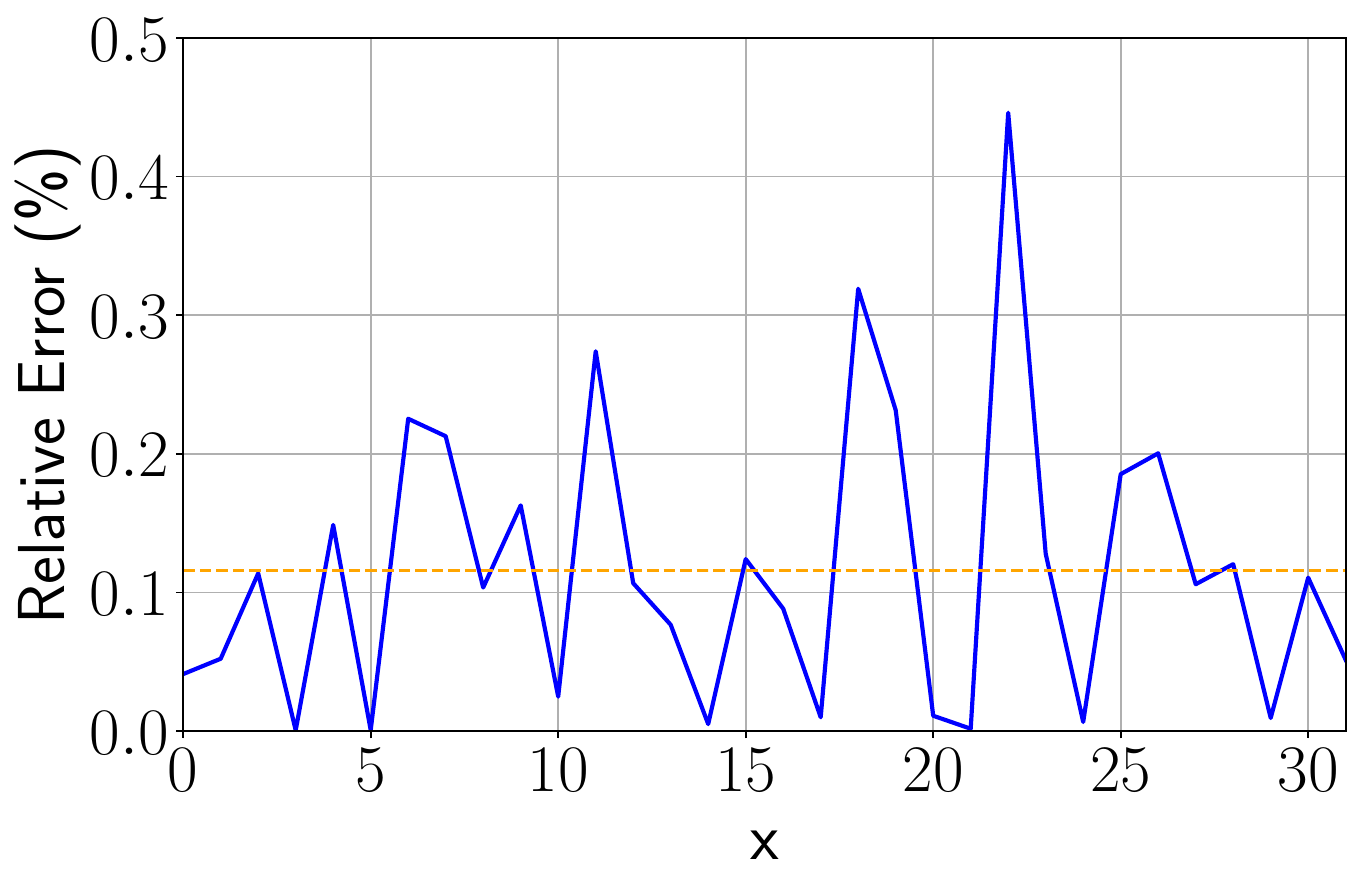}
        \caption{Relative error for 200 time steps.}
        \label{fig:D1Q3_200_time_steps_non_uniform_u_error}
    \end{subfigure}
    \\
    \begin{subfigure}{0.4\textwidth}
        \includegraphics[width=\textwidth]{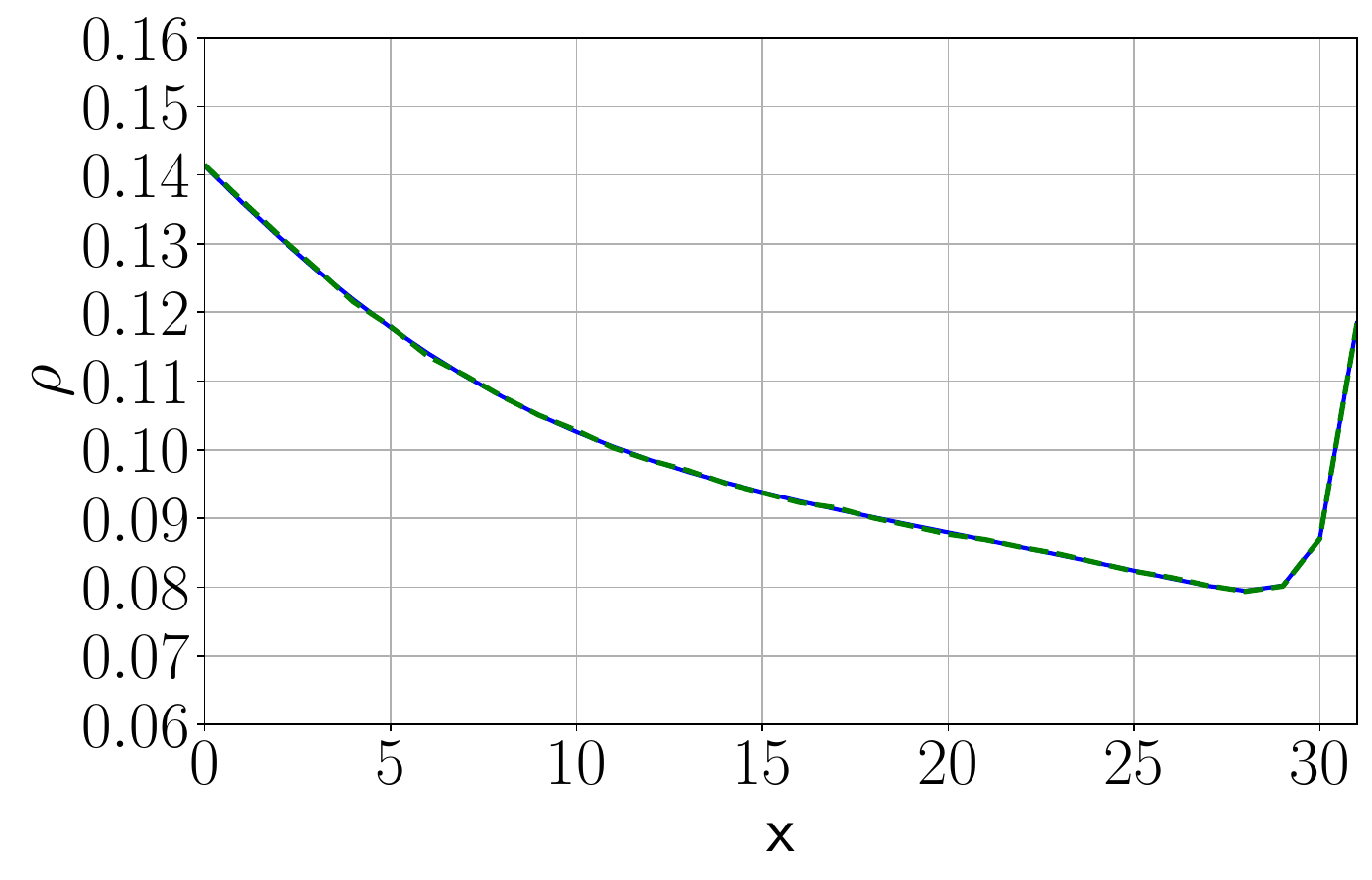}
        \caption{Result for 250 time steps.}
        \label{fig:D1Q3_250_time_steps_non_uniform_u}
    \end{subfigure}
    \hfill
    \begin{subfigure}{0.4\textwidth}
        \includegraphics[width=\textwidth]{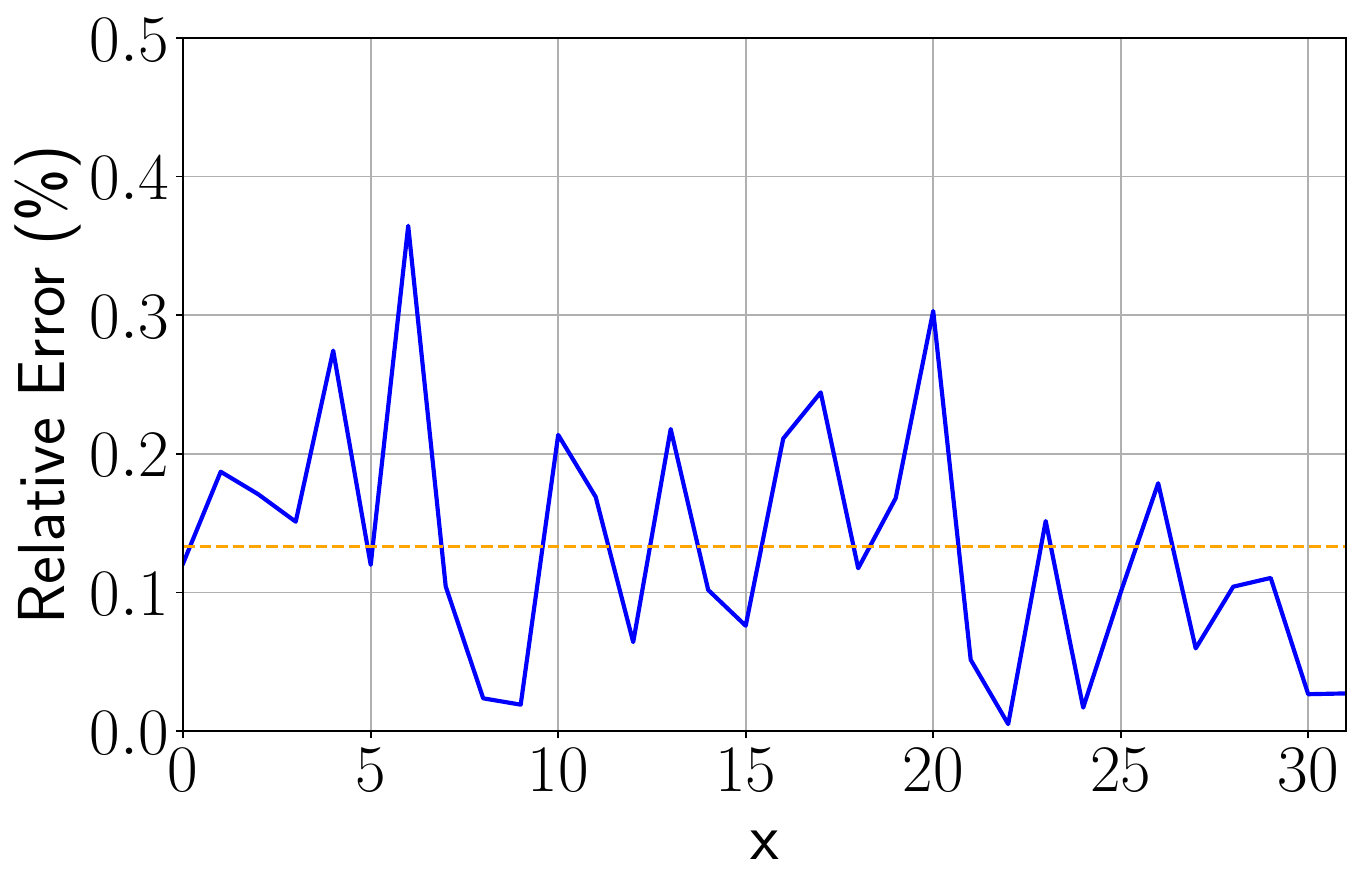}
        \caption{Relative error for 250 time steps.}
        \label{fig:D1Q3_250_time_steps_non_uniform_u_error}
    \end{subfigure}
    \caption{Results for QLBM and digital LBM using the D1Q3 velocity set for multiple time steps with $10^7$ shots and a non-uniform advection velocity. 
 In the left column, the digital result (\fullblack{}), and the QLBM result (\dashedgreen{}) are shown. In the right column, the relative error (\fullblue{}), and the MAPE (\dashedorange{}), are depicted.}
     \label{fig:D1Q3_multiple_t_10_7_shots}
\end{figure}

The results for $t= [50,150,200,250]$ time steps are presented in the left column of \cref{fig:D1Q3_multiple_t_10_7_shots}, while the corresponding relative errors at each lattice site are shown in the right column. 
The MAPE, depicted as a horizontal dashed line, remains nearly constant across all time steps. 
Variations in the local relative error are attributed to shot noise.

These findings suggest that, in the cases examined, once the number of shots is chosen to achieve a desired error threshold, maintaining this number of shots throughout multiple time steps remains sufficient to preserve the accuracy level.

\subsection{Hybrid quantum algorithm}

\begin{figure}
    \centering
    \begin{subfigure}{0.45\textwidth}
        \includegraphics[width=\textwidth]{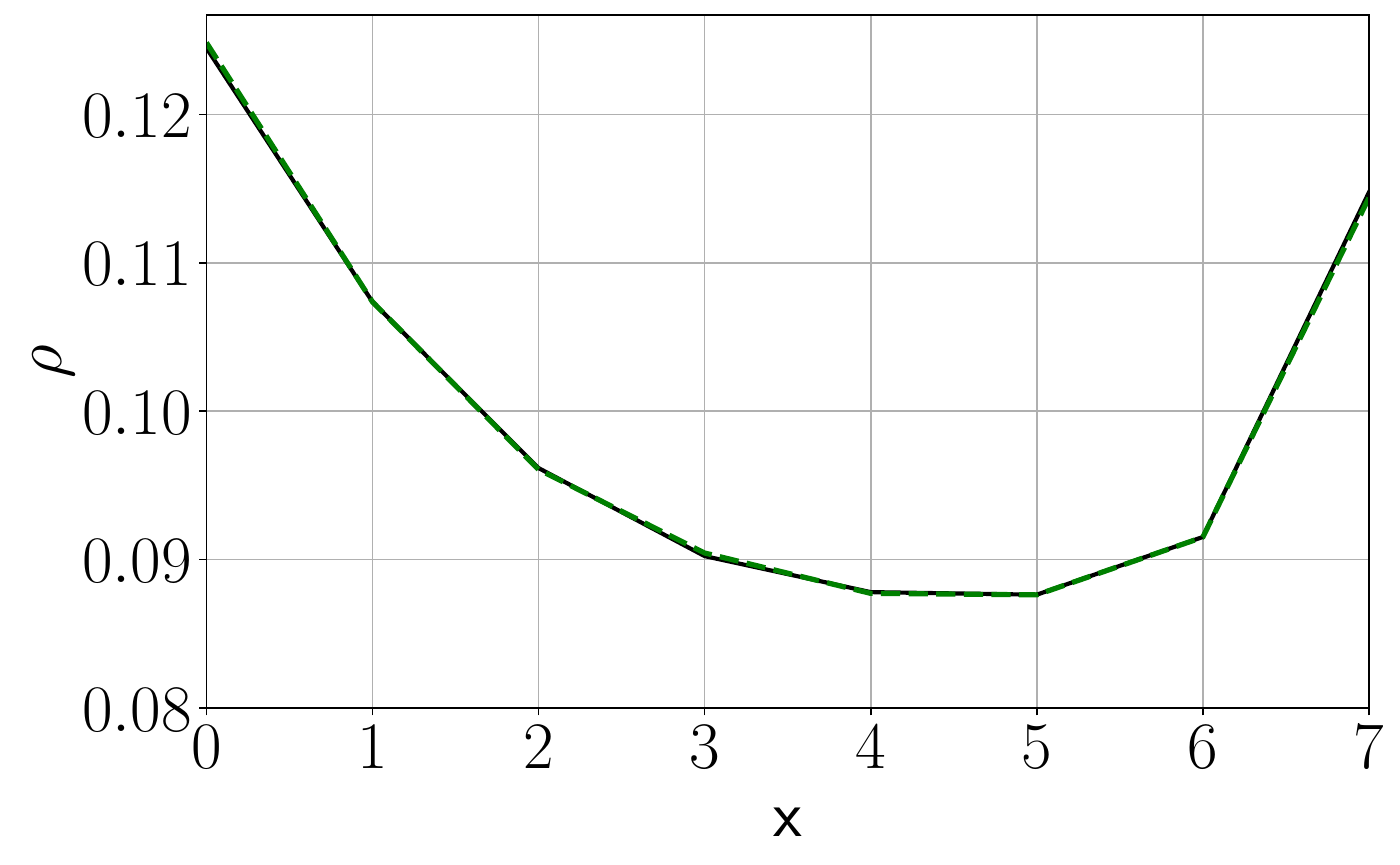}
        \caption{Result for 10 time steps of the hybrid quantum algorithm.}
        \label{fig:D1Q3_hybrid_10_time_steps}
    \end{subfigure}
    \hfill
    \begin{subfigure}{0.45\textwidth}
        \includegraphics[width=\textwidth]{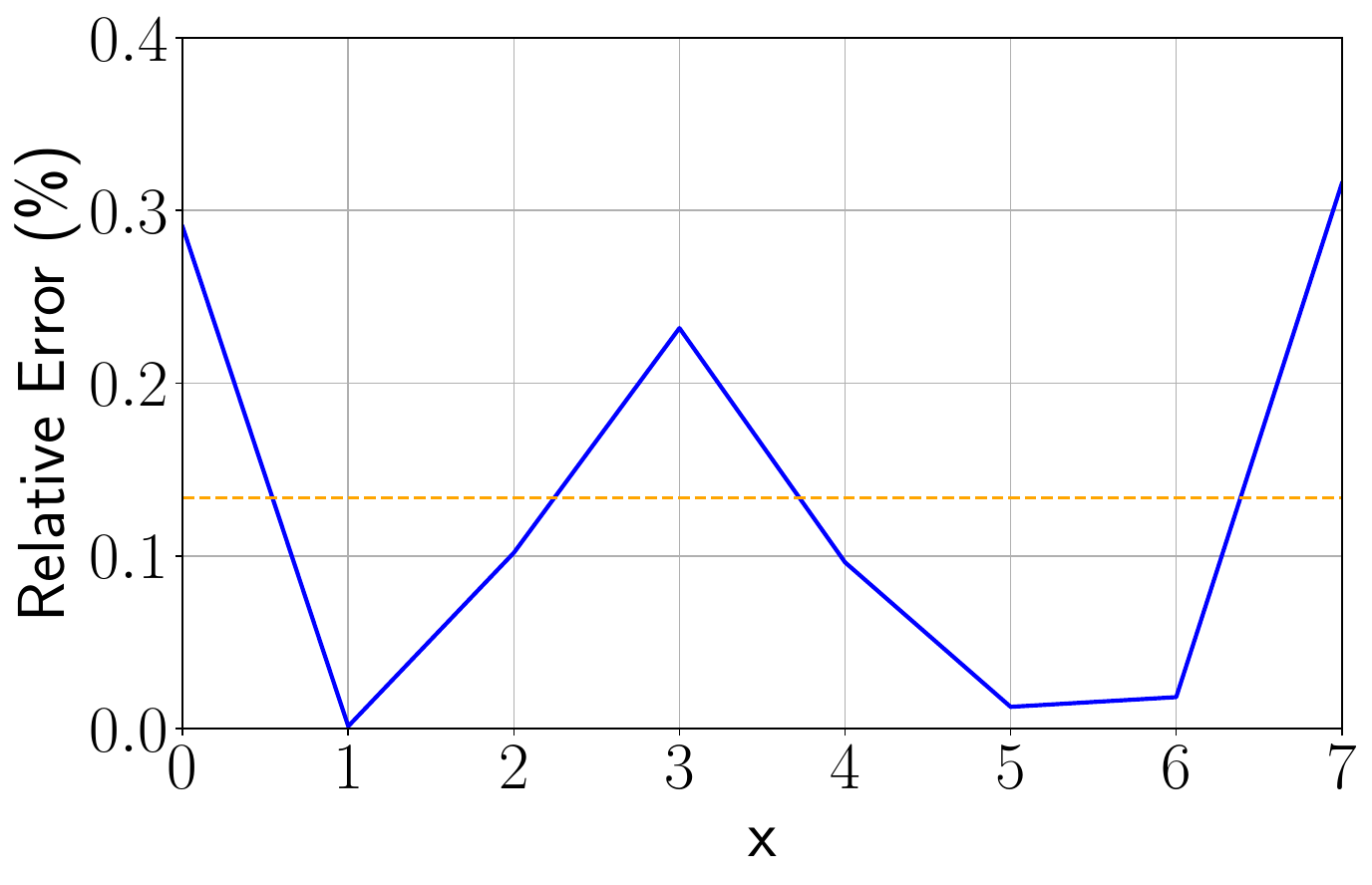}
        \caption{Relative error for 10 time steps of the hybrid quantum algorithm.}
        \label{fig:D1Q3_hybrid_10_time_steps_relative_error}
    \end{subfigure}
    \caption{Results and relative error for hybrid QLBM and digital LBM using the D1Q3 velocity set for 10 time steps with $10^6$ shots. 
 The digital result (\fullblack{}), and the hybrid QLBM result (\dashedgreen{}), are shown.
 The relative error (\fullblue{}), and the MAPE (\dashedorange{}), are depicted.}
    \label{fig:D1Q3_hybrid_quantum_algorithm}
\end{figure}

We consider a uniform initial density $\rho(x, t=0) = 0.1$ and a linearly increasing advection velocity field $u(x) = 0.1x + 0.1$, where $x \in [0,1]$ represents normalized spatial coordinates. 
The simulation is performed for 10 time steps using the hybrid quantum algorithm and $10^6$ shots on a domain with $N=8$ lattice cells, employing a D1Q3 velocity set.
In the digital preprocessing step, a weighted distribution is defined such that the outcome 0 is drawn in $2/3$ of all cases, and the outcome 1 is drawn in $1/3$ of all cases. 
These weights correspond to the D1Q3 velocity set, where $w_0 = 2/3$ and $w_1 + w_2 = 1/3$. A sequence of 10 random draws is then generated from this distribution, representing the computation of 10 time steps.
This procedure is repeated $10^6$ times, resulting in $10^6$ samples of sequences, each consisting of 10 independently drawn outcomes. 
These sequences serve as the blueprint for constructing the corresponding quantum circuits. No operation is applied for an outcome of 0, which corresponds to the identity operator. 
For an outcome of 1, the corresponding collision and streaming dynamics are implemented through the appropriate quantum circuit.
Each quantum circuit is executed with one shot.
The results are presented in \cref{fig:D1Q3_hybrid_10_time_steps}, with the corresponding relative error at each lattice site and the MAPE shown in \cref{fig:D1Q3_hybrid_10_time_steps_relative_error}. 
The hybrid QLBM demonstrates good agreement with the digital LBM, as the minor relative errors indicate. 
Therefore, we conclude that the hybrid quantum algorithm accurately reproduces the reference simulation.

\section{Conclusion}\label{sec.5}

The primary contribution of this work lies in developing and applying dynamic circuits for a quantum-biased linearized LBM collision, enabling efficient and reliable time stepping in QLBM. 
Dynamic circuits leverage mid-circuit measurements and digital processing during quantum circuit execution to determine subsequent unitary operations dynamically. 
This methodology significantly reduces qubit overhead and the number of qubits on which operators are applied, simplifying the quantum algorithm while enhancing its efficiency.

Moreover, we provide further algorithmic optimization for the collision and streaming steps compared to prior implementations. 
By employing the square root of equilibrium distribution functions, the collision step is efficiently expressed using unitary operators, implemented with \texttt{UCRY} gates. 
This reduces the complexity of digital preprocessing, as a certain direction of the discretized distribution and its opposite are inherently linked in the quantum algorithm.
This unitary approach ensures conservation properties, avoiding issues such as mass loss, which are prevalent in previous algorithms relying on probabilistic collision operator methods \citep{wawrzyniak2025quantum}.

The advantages of the proposed algorithm are especially prominent for multi-time-step computations. 
Unlike previous quantum algorithms that utilize linear combinations of unitaries for non-unitary classical collision operators, which rely on probabilistic application and prevent reliable time stepping \citep{budinski2021quantum,ljubomir2022quantum,wawrzyniak2025quantum}, the dynamic circuit approach allows for stable propagation through multiple time steps. 
After initialization, the same dynamic circuit is repeated for each time step, eliminating the need for total state measurements and reinitialization. 
In our tests, we have successfully computed up to 250 time steps without requiring excessive shot numbers, demonstrating the robustness of the approach.

In previous work, reinitialization introduces high sensitivity to shot noise, necessitating an impractically large number of shots per time step to maintain noise-free solutions \citep{wawrzyniak2025quantum}. 
Without sufficient shot count, these methods exhibit significant errors, such as mass loss and divergence from the expected solution trajectory. 
The proposed unitary collision operator inherently avoids such issues, resulting in a more stable and accurate overall framework.

We also introduce a novel streaming implementation within the dynamic circuit QLBM algorithm. 
Unlike previous algorithms that apply streaming operations to all distribution functions, irrespective of measurement outcomes, our approach applies the streaming operator only to active distribution functions as determined by mid-circuit measurements. 
For instance, in a D2Q9 velocity set, prior methods would apply nine streaming circuits per time step, whereas our algorithm requires only one, largely reducing the computational cost.

The hybrid digital-quantum nature of our QLBM algorithm further enhances its practicality. 
Using a classical random sampling process to emulate the quantum circuit structure, we minimize the number of mid-circuit measurements while maintaining efficiency. 
Resting distribution functions are integrated with a simple identity operator, reducing unnecessary operations.

In summary, the proposed dynamic circuit-based QLBM algorithm represents a step forward in quantum Lattice-Boltzmann methods, offering improved scalability, reduced complexity, and enhanced stability for multi-time-step simulations. 
\section*{Acknowledgments}
The authors gratefully acknowledge funding for this research from Altair Engineering Inc.

\appendix

\section{Quantum circuits for the D2Q9 velocity set}\label{App.1}

\begin{figure}
    \centering
    \begin{subfigure}{0.45\textwidth}
        \includegraphics[width=\textwidth]{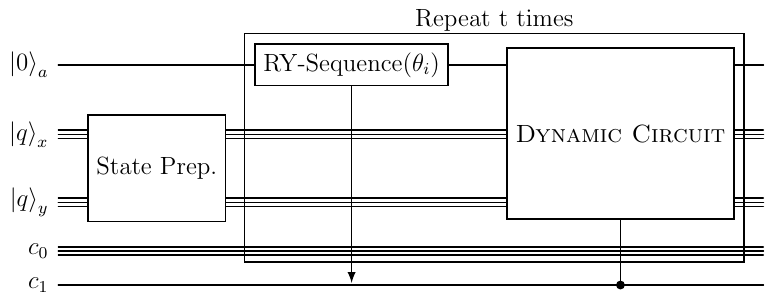}
        \caption{The general quantum circuit for the D2Q9 QLBM.}
        \label{fig:full_circuit_D2Q9}
    \end{subfigure}
    \hfill
    \begin{subfigure}{0.45\textwidth}
        \includegraphics[width=\textwidth]{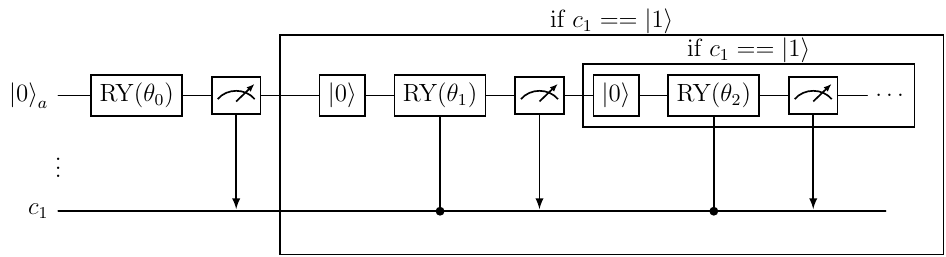}
        \caption{The RY sequence circuit for the D2Q9 QLBM.}
        \label{fig:ry_sequence_circuit_D2Q9}
    \end{subfigure}
    \hfill
    \begin{subfigure}{0.45\textwidth}
        \includegraphics[width=\textwidth]{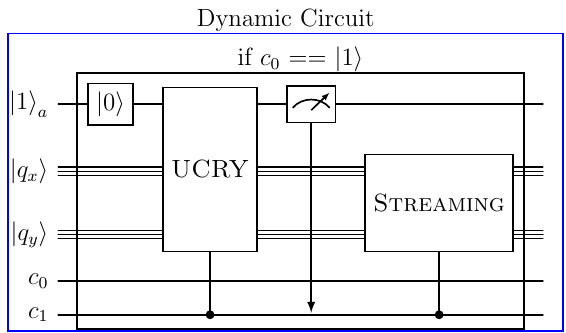}
        \caption{Dynamic circuit for the D2Q9 QLBM.}
        \label{fig:dynamic_circuit_D2Q9}
    \end{subfigure}
    \hfill
    \begin{subfigure}{0.45\textwidth}
        \includegraphics[width=\textwidth]{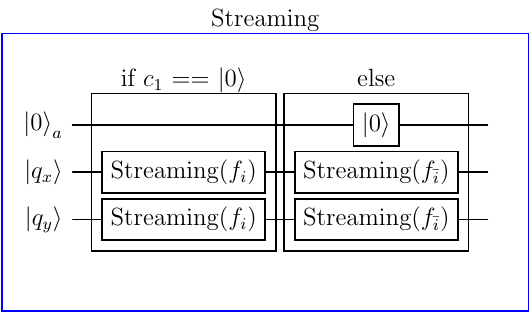}
        \caption{Streaming circuit for the D2Q9 QLBM.}
        \label{fig:streaming_D2Q9}
    \end{subfigure}
    \caption{All quantum circuits for the dynamic circuit implementation of the D2Q9 QLBM.}
    \label{fig:all_circuits_D2Q9}
\end{figure}

In this section, we present the quantum circuit for implementing the QLBM with the D2Q9 velocity set. 
\Cref{fig:full_circuit_D2Q9} illustrates the quantum circuit, utilizing general operators for the setup. 
The RY sequence is executed with the parameter vector
\begin{equation}
    \theta = [2\arccos(\sqrt{4/9}), 2\arccos(\sqrt{2/5}), 2\arccos(\sqrt{2/3}), 2\arccos(\sqrt{1/2})]^T,     
\end{equation}
where each subscript in the circuit corresponds to an entry in this vector. 
This vector is designed to ensure that the amplitudes of the ancillary qubit superpositions in the applied sequence align with the LBM weights. This relationship is demonstrated as
\begin{equation}
    RY(\theta_0)\ket{0} = \sqrt{4/9}\ket{0} + \sqrt{5/9}\ket{1}.
\end{equation}
Thus, the measurement probabilities are $p(\ket{0})=4/9$ and $p(\ket{1})=5/9$, corresponding to $w_0$ and $\sum_{i=1}^8 w_i$, respectively. 
The state following the first RY gate is denoted as $\ket{\Psi_0} = \ket{0}$. In cases where the measurement outcome is $\ket{1}$, resetting and applying $RY(\theta_1)$ produces
\begin{equation}
    RY(\theta_1)\ket{0} = \sqrt{2/5}\ket{0} + \sqrt{3/5}\ket{1}.    
\end{equation}
The measurement probabilities are $p(\ket{0})=2/5$ and $p(\ket{1})=3/5$. 
For the quantum algorithm, $p(\ket{0})$ corresponds to $w_1 + w_3 = 2/9$. This is verified using the density operator formalism $\sigma$ after the second RY gate
\begin{equation}
    \sigma = p_1 \ket{\Psi_0}\bra{\Psi_0} + p_2 \ket{\Psi_1}\bra{\Psi_1}.
\end{equation}
The probability of measuring $\ket{0}$ after the second RY gate is $p_3(\ket{0})=\frac{5}{9} \cdot \frac{2}{5} = \frac{2}{9}$, matching $w_1 + w_3$. 
This process is iteratively executed, with each step associating the measurement outcomes with specific lattice directions and their opposites. 
Once a measurement yields either $\ket{0}$ or $\ket{1}$ following the final RY-gate application, the sequence terminates, and the corresponding dynamic circuit is applied to the quantum state.
The dynamic circuit incorporates a collision block implemented through a UCRY operator and a mid-circuit measurement that conditionally activates the specific streaming algorithm, dynamically adjusting based on the measurement outcomes.
The dynamic circuit is shown in \cref{fig:dynamic_circuit_D2Q9} and the streaming circuit is shown in \cref{fig:streaming_D2Q9}.

\bibliographystyle{elsarticle-num} 
\bibliography{refs.bib}





\end{document}